\documentclass[aps,prl,nobalancelastpage,twocolumn,superscriptaddress,reprint]{revtex4-2}
\usepackage{multirow}
\usepackage{hyperref}
\hypersetup{
    colorlinks=true,
    linkcolor=blue,
    filecolor=gray,      
    urlcolor=blue,
    citecolor=blue,
}
\usepackage{graphicx}
\usepackage{rotating}

\usepackage{bm, physics}
\usepackage{tabularx}
\usepackage{booktabs}
\usepackage{amsfonts, amssymb, amsmath}

\begin{document}

\title{Chemical design of monolayer altermagnets}

\affiliation{Department of Physics, The Hong Kong University of Science and Technology, Clear Water Bay, Hong Kong, China}

\author{Runzhang \surname{Xu}}
\email{xurz@ust.hk}
\author{Yifan \surname{Gao}}
%
%
%
%
\author{Junwei \surname{Liu}}
\email{liuj@ust.hk}


\begin{abstract}

The crystal-symmetry-paired spin-momentum locking (CSML) arisen from the intrinsic crystal symmetry connecting different magnetic sublattices in altermagnets enables many exotic spintronics properties such as unconventional piezomagnetism and noncollinear spin current.
However, the shortage of monolayer altermagnets restricts further exploration of dimensionally confined phenomena and applications of nanostructured devices.
Here, we propose general chemical design principles inspired by sublattice symmetry of layered altermagnet V$_2$(Se,Te)$_2$O through symmetry-preserving structural modification and valence-adaptive chemical substitutions.
In total, we construct 2600 candidates across four structural frameworks, M$_2$A$_2$B$_{1,0}$ and their Janus derivatives.
High-throughput calculations identify 670 potential altermagnets with N\'eel-ordered ground states, among which 91 ones exhibiting CSML Dirac cones that enable spin-polarized ultra-fast transport.
These materials also feature different ground-state magnetic orderings and
demonstrate diverse electronic behaviors, ranging from semiconductors, metals, half-metals, to Dirac semimetals.
This work not only reveals abundant monolayer altermagnets, but also establishes a rational principle for their design,
opening gates for exploration of confined magnetism and spintronics in atomically thin systems.
~

~

\end{abstract}

\maketitle
\section{Introduction}

\begin{figure*}[t]
   \centering
   \includegraphics[width=0.9\textwidth]{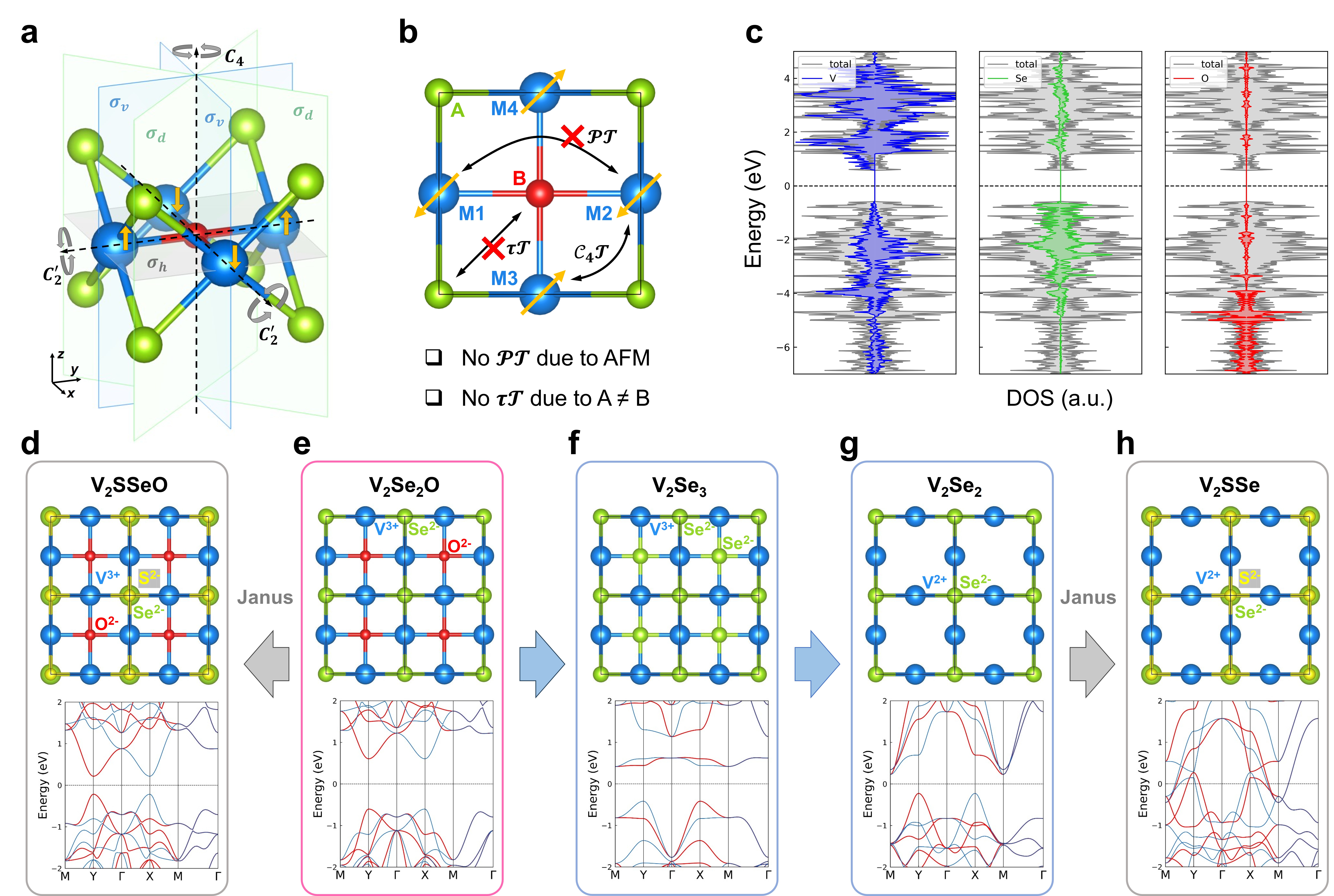}
   \caption{(\textbf{a}) Schematic crystal structure of V$_2$Se$_2$O monolayers. Blue, green, and red spheres stand for V, Se, and O. Its mirror and rotation symmetries are presented by transparent colored planes and combinations of black dashed and gray rotating arrows, with respective notations shown beside. The N\'eel AFM-ordered spins on V are denoted by yellow arrows on blue spheres. (\textbf{b}) Structural framework M$_2$A$_2$B as design basis. Atomic sites M, A, and B resemble the in-plane locations of metal V, vertical-dimmer Se-Se, and single-ion O in V$_2$Se$_2$O, respectively; the symmetry requirements of altermagnetism is also marked inset. (\textbf{c}) Density of states of V$_2$Se$_2$O (gray) and its projected ones onto V, Se, and O ions (blue, green, and red, respectively). (\textbf{d-h}) Top view of crystal structures (top panel) of Janus V$_2$SSeO, V$_2$Se$_2$O, V$_2$Se$_3$, V$_2$Se$_2$, and Janus V$_2$SSe, as well as their band structures (bottom panel) under N\'eel-AFM order. The in-between arrows denote the symmetry-preserving structural modification by site-B engineering (blue) and Janus structuring (gray) in our design principles. Valence of each constituent elements is explicitly marked in crystal structures, and top S and bottom Se ions in \textbf{c} and \textbf{g} are distinguished by small yellow and large green spheres. }
   \label{fig:1}
\end{figure*}

Altermagnets simultaneously feature both spin-splitting band structures with crystal-symmetry-paired spin-momentum locking (CSML) in momentum space
and antiferromagnetic (AFM) ordering in real space due to the intrinsic crystal symmetries
that map magnetic sublattices of opposite spins to each other \cite{CJWu2009,naka2019spin,Hayami2019,sciadv_altermag,YLDong2019,ma2021multifunctional,mejkal2022,altermag_prx}.
These features give rise to unique properties including unconventional spiezomagnetism \cite{CJWu2009, ma2021multifunctional} and non-collinear spin current \cite{CJWu2009, naka2019spin,ma2021multifunctional}, which not only advances our understanding of unconventional magnetism but also promises many novel applications \cite{song2025altermagnets,Bai2024,Fender2025}. Several candidates for real materials have been confirmed in experiments \cite{Bai2024,song2025altermagnets}, 
while most of them are bulk crystals or thin films,
with the only layered example being V$_2$(Se,Te)$_2$O \cite{ma2021multifunctional,Zhang2025,Jiang2025}.
The lack of two-dimensional (2D), in particular monolayer, altermagnetic (AM) materials strongly hinders the further investigation of novel phenomena and exotic physics due to the reduced dimensionality and strong correlations as well as their susceptibility to external modulations (\textit{e.g.}, strain, doping, and/or electric gating), proximity effects from superconducting or topological phases, and facile integration with existing nano-structured devices \cite{Tanaka2007,Xu2014}.

A highly effective approach to discover new materials of novel properties relies on empirically validated chemical design principles, leveraging insights from existing materials to guide the exploration of new ones.
Two prominent examples illustrate this approach include the comprehensive exploration of MoS$_2$-type 2D transition-metal dichalcogenides (TMDCs) \cite{Wilson1975,mak2010,Qian2014,Mak2016,manzeli20172d} and the discovery of the intrinsic quantum anomalous Hall (QAH) insulator MnBi$_2$Te$_4$ \cite{Li2019,Gong_2019,Zhang2019,He2019,Deng2020}.
TMDCs exemplify the success of chemical design due to their vast configuration space arising from diverse chemical elements and structural phases. Their electronic properties are profoundly influenced by their structure: the 2H phase typically exhibits a semiconducting state with spin-valley locking \cite{Xiao2012}, while the 1T phase often displays metallic behavior, regardless of its constituent elements. Manipulating the 1T phase through element substitution can lead to more intriguing outcomes. 
For instance, replacing certain elements may induce superconductivity through the melting of intrinsic charge density waves (\textit{e.g.}, TaS$_2$ \cite{Sipos2008,NavarroMoratalla2016} and TiSe$_2$ \cite{Kusmartseva2009,Calandra2011,Joe2014}) or cause electronic instability and Jahn-Teller distortion, leading to a transition to the 1T' phase (or even Td phase) and hosting quantum spin Hall (QSH) effects (\textit{e.g.}, (Mo,W)Te$_2$ \cite{Ali2014,Qian2014,LiuJW2016,Fei2018,Sharma2019}). Further, Janus structuring, which breaks inversion symmetry, introduces the Stark effect and Rashba spin splitting (\textit{e.g.}, MoSSe and WSSe \cite{Li2017,Zhou2019wsse}).
The discovery of MnBi$_2$Te$_4$ highlights another example of chemical design. The pursuit of QAH insulators \cite{Liu2016,Chang2023} initially involved extrinsically doping magnetic ions (Cr or V) into the QSH insulator (Sb,Bi)$_2$Se$_3$ \cite{Yu2010,Chang2013}. Later advancements led to the intrinsic formation of MnBi$_2$Te$_4$ by depositing MnTe onto the topological insulator Bi$_2$Te$_3$ \cite{Li2019,Gong_2019,Zhang2019,He2019,Deng2020}.
This chemical design approach also extends to many other material classes and novel properties, such as SnTe- and Sr$_3$PbO-type topological crystalline insulators \cite{Hsieh2012,Tanaka2012,Qian2014,Chang2016,Huang2019,Lam2023}, and WTe$_2$- and TaIrTe$_4$-type QSH insulators \cite{Tang2017,Shi2019,Garcia2020,LiuJW2016,Tang2024}.

In this work, inspired by the V$_2$(Se,Te)$_2$O structure and its symmetry-preserving variants (both modified and Janus structured), along with the potential for valence-adaptable element substitution at all atomic sites, we propose systematic chemical design principles for monolayer altermagnets.
In total, 2600 material candidates are rationally designed in four structural frameworks,
including 880, 220, 1200, and 300 for M$_2$A$_2$B and M$_2$A$_2$B, and Janus structured M$_2$AA$'$B and M$_2$AA$'$, respectively,
where M is a transition metal,
A/A$'$ comes from either pnictogens, chalcogens or halogens,
and B is from chalcogens.
The substitution elements are chosen with consideration of sustaining strong covalent bonding,
compensating element valence,
and avoiding fractional valence.
Our high-throughput first-principles calculations of all 2600 candidates cover a broad range of magnetic orderings at ground state,
including AM, non-magnetic (NM), ferromagnetic (FM), and stripe-AFM,
as well as electronic structures under ground-state orderings,
namely semiconducting, metallic, half-metallic, and Dirac-cone semi-metallic.
In the calculations,
plenty of altermagnets (143, 81, 335, and 111, respectively) are identified from candidates of all four structural frameworks. Surprisingly,
some of them (11, 17, 39, and 24, respectively) can simultaneously host CSML Dirac cones near the Fermi surface,
suggesting potential anisotropic and spin-selective carrier transport at ultra-high speed for future ultra-fast spintronics applications.
Our designs and results provide guidance for future investigations of monolayer altermagnets and their distinct electronic structures and properties.

\section{Results and Discussion}
\subsection{Symmetry and Design Principles}


The 2D crystal structure of layered V$_2$(Se,Te)$_2$O \cite{ma2021multifunctional} belongs to point group $D_{4h}$ and space group $P4/mmm$, as depicted in Fig. 1a.
Its square lattice is symmetric under rotations of four-fold principal $\mathcal{C}_4$ and two-fold non-principal $\mathcal{C}_2'$,
mirrors of horizontal $\sigma_h$, vertical $\sigma_v$, and dihedral $\sigma_d$,
as well as multiple centers of inversion $\mathcal{P}$.
The two spin-antiparallel magnetic sublattices (Fig. S1) can only be connected by the $\mathcal{C}_4$ rotation and $\sigma_d$ mirror symmetry,
rather than the $\mathcal{P}$ or any half translations $\tau$,
which leads to the spin-splitting band structures even without spin-orbit coupling
and enforces the crystal-symmetry paired spin-momentum locking (CSML).
This is the direct consequence of identical spins at site M1 and M2 (also, M3 and M4),
and different non-metal chemical species at site A and B (vertical-dimmer Se and single-ion O),
as shown in Fig. 1b.
As long as the A $\ne$ B condition and checkboard-type AFM ordering withstand,
the critical breaking of $\mathcal{P}\mathcal{T}$ and $\tau\mathcal{T}$ is preserved,
and the atomic configurations of site A and B can be changed by
either modification or even site subtraction of site B,
or Janus structure of two vertically-stacked elements at site A
(\textit{i.e.} double A becoming Janus AA$'$).
It is worth mentioning that the vertical translations are absent in 2D regime and
hence the spin-antiparallel magnetic sublattices can only be connected by $\mathcal{C}_4$ (or other even-fold) but not $\mathcal{C}_3$.
Therefore, preserving the critical symmetry breaking
and suitable magnetic structure is crucial to realize monolayer altermagnets.

Taking advantage of the critical symmetry and structural framework of synthesized V$_2$(Se,Te)$_2$O (M$_2$A$_2$B in Fig. 1b),
we propose a systematic design approach for monolayer altermagnets through site modification and subtraction at the B site, and Janus structures at the A site.
The effectiveness of this design approach is exemplified by monolayer V$_2$Se$_2$O and its derivatives as shown in Fig. 1.
Projected density of states (Fig. 1c) reveal that the majority of oxygen (O) electronic states at site B reside deep in the valence regime, with negligible influence on states near the Fermi level and thus minimal impact on electronic properties and magnetic orderings. 
Either replacing O with Se or removing it does not affect any symmetry and indeed maintains the semiconducting nature and AM order, as demonstrated in Fig. 1e-g.
The in-gap flat bands (Fig. 1f) and the conduction valley shift (Fig. 1g) can be attributed to enhanced correlation from the more active Se outer shells and reconstructed hybridization resulting from the removal of O $p_x/p_y$ orbitals.
In addition, introducing a Janus structure, which still preserves the critical symmetry breaking, can introduce additional symmetry breaking and novel characteristics.
To ensure stability, we substitute one of the vertically-stacked Se atoms at site A with congeneric S, resulting in AM band structures with narrowing and even closing gaps in V$_2$SSeO and V$_2$SSe (Fig. 1c and h) due to (vertical) symmetry breaking-induced Stark effect.
Inspired by these examples, we construct four structural frameworks, M$_2$A$_2$B and M$_2$A$_2$, and their Janus derivatives M$_2$AA$'$B and M$_2$AA$'$, for further design.

We next implement symmetry-allowed substitution of chemical elements at M, A/A$'$, and B atomic sites within four design frameworks,
with element selection aiming to maintain key chemical properties.
The non-metal elements at site A and/or B are chosen from group VA-VIIA (in periodic table) to ensure strong covalent bonding analogous to that in V$_2$(Se,Te)$_2$O.
While the metal elements at site M are chosen from transition metals (group IIIB-IIB in periodic table), due to their potential of hosting spin polarization and multiple oxidation states derived from partially filled $d$ orbitals.
We neglect very heavy elements and focus on ones within period 5 to match the common research interest.
Consequently, the substituting elements encompass the first 20 transition metals (Sc-Cd) and the first 4 pnictogens (N-As), chalcogens (O-Te), and halogens (F-I), with metallic Sb being subtracted.
To ensure the stability/metastability of the design candidates after substitution,
the closed-shell electronic configuration (\textit{e.g.}, N$^{3-}$:[Ne], O$^{2-}$:[Ne], and F$^{1-}$:[Ne]) is required for non-metal elements,
and the fractional or mixed valences are also avoided by restricting elements at site B to have even-integer valences (\textit{i.e.}, only chalcogens).
Notably, the stability of transition metals is governed less by closed $d$ shell, but rather by ligand field effects and specific $d$-shell fillings (\textit{i.e.}, empty $d^0$ for early transition metals and half/full $d^5$/$d^{10}$ for late ones).
Applying these rules and leveraging the multiple oxidation states enabled by transition-metal $d$ shells,
we design all feasible valence combinations (or "recipes") for the substituting elements in four aforementioned frameworks (Table I).
Therefore, a total of 2600 potential monolayer altermagnet candidates are generated by our design for high-throughput investigations, which include 880 for M$_2$A$_2$B, 220 for M$_2$A$_2$, 1200 for Janus M$_2$AA$'$B, and 300 for Janus M$_2$AA$'$.

\begin{table}[]
	\centering
    \renewcommand\arraystretch{1.4}
    \setlength{\tabcolsep}{8pt}
    \caption{Possible valence recipes for substituting elements at site M, A, A$'$, and B of four AM structural frameworks within the chemical design.}
	\begin{tabular}{ccccc}
		\hline
        Design framework     & M  & A  & A$'$  & B  \\ \hline
		M$_2$A$_2$B          & +2        & $-$1      & --           & $-$2      \\
                             & +3        & $-$2      & --           & $-$2      \\
                             & +4        & $-$3      & --           & $-$2      \\ \hline
		M$_2$A$_2$           & +1        & $-$1      & --           & --        \\
                             & +2        & $-$2      & --           & --        \\
                             & +3        & $-$3      & --           & --        \\ \hline
        M$_2$AA$'$B          & +2        & $-$1      & $-$1         & $-$2      \\
        (Janus)              & +3        & $-$2      & $-$2         & $-$2      \\
                             & +4        & $-$3      & $-$3         & $-$2      \\ \hline
		M$_2$AA$'$           & +1        & $-$1      & $-$1         & --        \\
        (Janus)              & +2        & $-$2      & $-$2         & --        \\
                             & +3        & $-$3      & $-$3         & --        \\ \hline
	\end{tabular}
	\label{tab:table_1}
\end{table}

\begin{figure*}[t]
   \centering
   \includegraphics[width=1.0\textwidth]{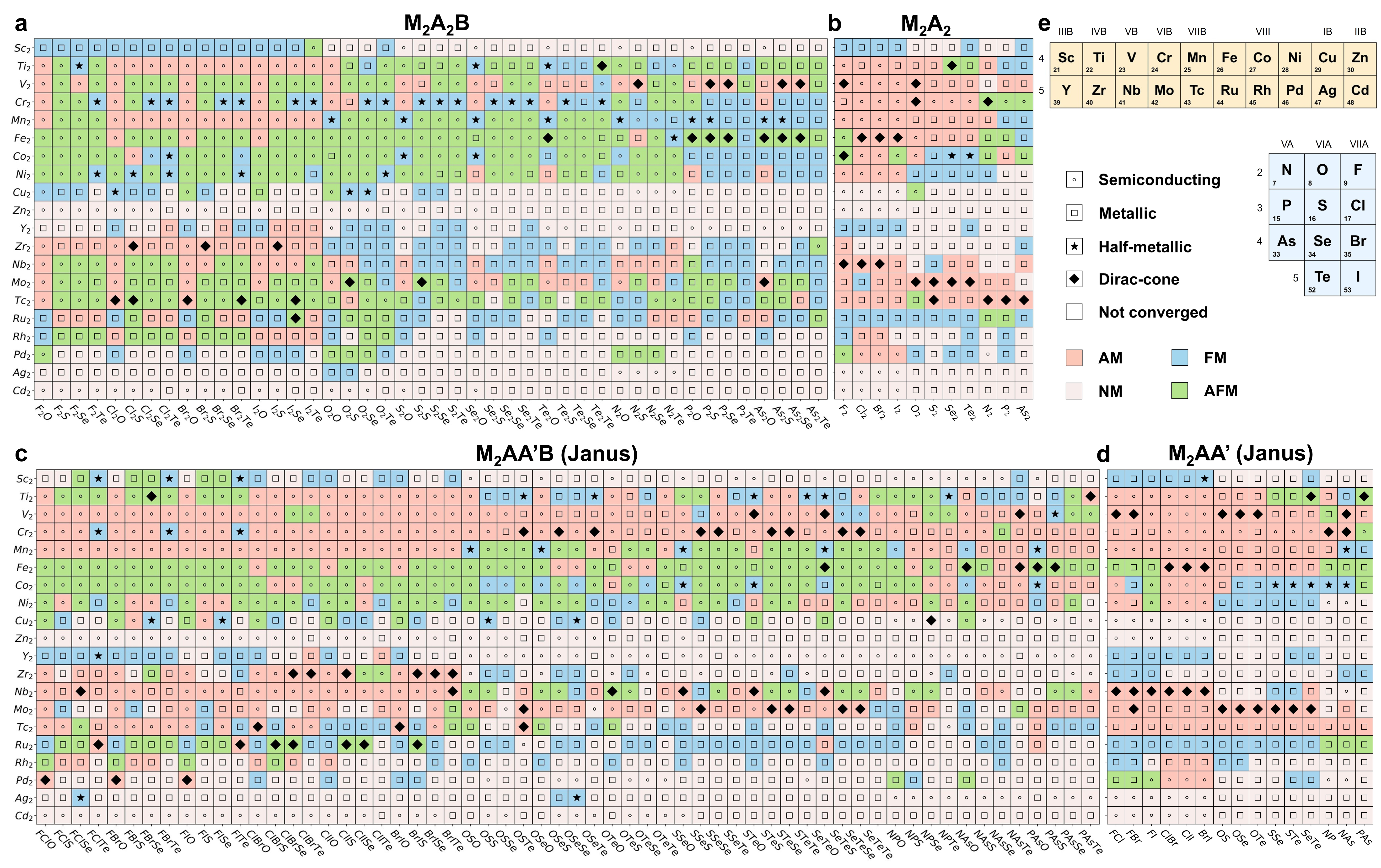}
   \caption{Ground-state magnetic orderings and electronic band structures of designed candidates from framework (\textbf{a}) M$_2$A$_2$B, (\textbf{b}) M$_2$A$_2$, (\textbf{c}) Janus M$_2$AA$'$B, and (\textbf{d}) Janus M$_2$AA$'$. The vertical and horizontal axes denote the metal and non-metal parts of the chemical formula. The horizontal altermagnetism-absent metal belts, the vertical dependence of AM order on non-metal elements, and the strong relation between Dirac cones and AM order are all clearly demonstrated. (\textbf{e}) Schematic periodic table showing the transition-metal and non-metal elements considered in our design. }
   \label{fig:2}
\end{figure*}

\begin{figure*}[t]
   \centering
   \includegraphics[width=0.7\textwidth]{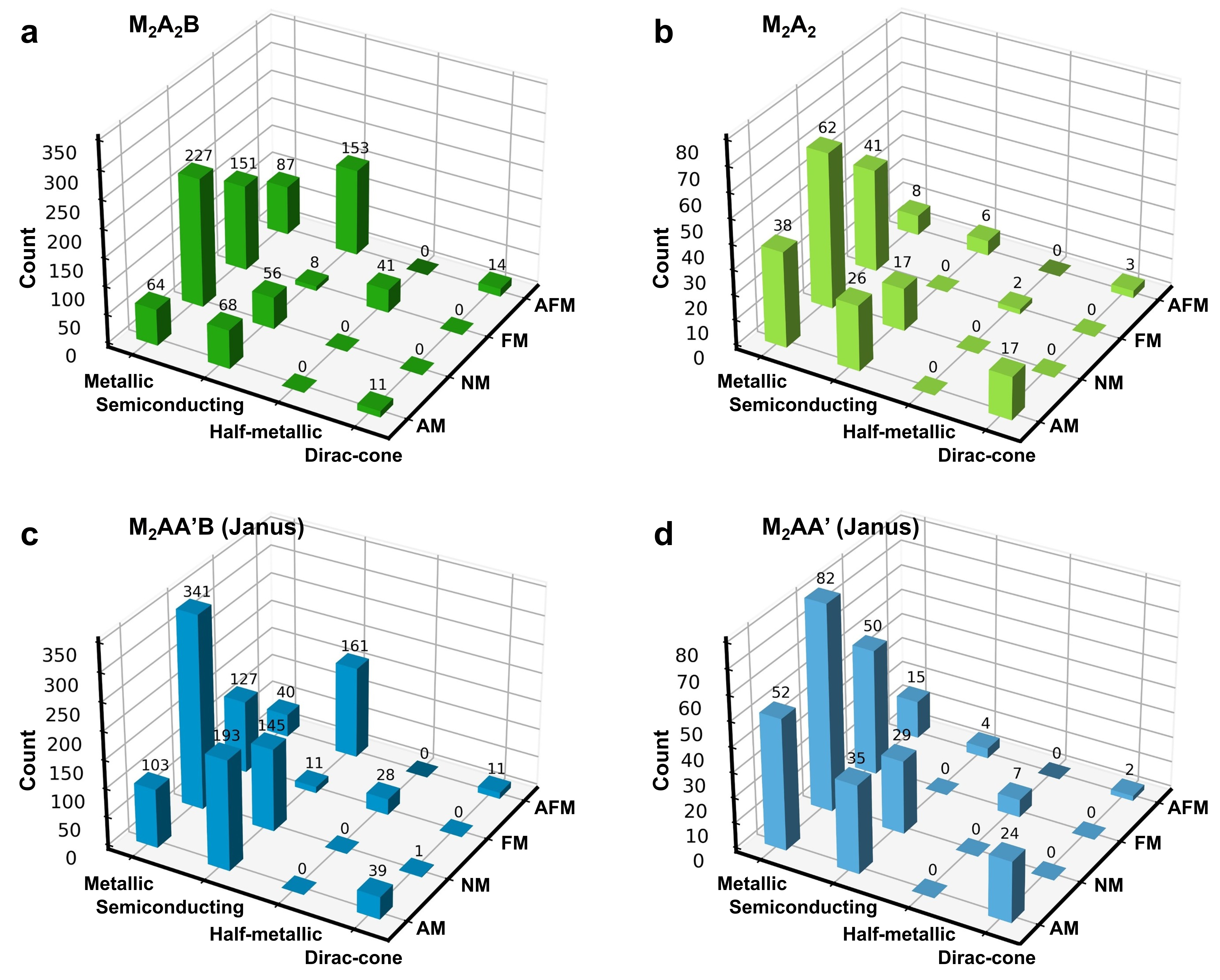}
   \caption{Statistical count of candidates with four different ground-state magnetic orderings and four distinct types of band structures for material design framework (\textbf{a}) M$_2$A$_2$B, (\textbf{b}) M$_2$A$_2$, (\textbf{c}) Janus M$_2$AA$'$B, and (\textbf{d}) Janus M$_2$AA$'$. The specific number for each pair of magnetic orderings and band-structure types is explicitly marked on top. }
   \label{fig:3}
\end{figure*}

\subsection{High-Throughput Calculations}

The magnetic orderings and their corresponding electronic band structures for all 2600 candidate materials are evaluated and obtained using high-throughput first-principles calculations.
Four magnetic orderings (or configurations) are considered in calculations (Fig. S2), namely AM (equivalently, N\'eel-AFM), stripe-AFM (AFM in short), ferromagnetic (FM), and non-magnetic (NM), and the lowest-energy one is identified as the ground state.
Electronic band structures are calculated self-consistently based on these ground-state magnetic orderings.

Our calculations identify 143 (16.3\% of 880), 81 (36.8\% of 220), 335 (27.9\% of 1200), and 111 (37.0\% of 300) potential monolayer altermagnets within the four respective design frameworks,
yielding a total of 670 taking up 25.8\% of the whole 2600.
And a diverse range of electronic properties, including semiconducting, metallic, half-metallic, and notably, Dirac-cone semimetallic, are also revealed.
The finding that a quarter of the designed candidates are indeed the desired altermagnets underscores the rationality and effectiveness of our design principles,
as well as contributes a substantial number of AM candidate materials for future experimental synthesis and investigation.
More importantly, among these identified altermagnets,
11, 17, 39, and 24 candidates from the respective frameworks are also found to simultaneously host CSML Dirac-cone energy bands near the Fermi level.
The ground-state magnetic orders and electronic band structures of all 2600 designed candidates are visually summarized in Fig. 2 using color fillings and centered symbols within mosaic squares (with corresponding statistical counts presented in Fig. 3),
consistent with findings from the few examples reported previously \cite{Chen2023,Zou2024}.

Our results clearly demonstrate the dependence of AM ordering on the constituent metal and non-metal elements.
In all four frameworks (Fig. 2), the emergence of AM order shows a distinct dependence on the transition-metal constituents,
leading to the formation of horizontal altermagnetism-absent belts of metals,
regardless of the non-metal constituents.
These belts span over 6 transition metals (Sc, Cu, Zn, and Pd-Cd) in M$_2$A$_2$B (Fig. 2a) and 7 (Pb replaced by Y and Ru) in M$_2$A$_2$ (Fig. 2b),
while changing significantly in Janus M$_2$AA$'$B and M$_2$AA$'$ (Fig. 2c and d) to comprise 4 (Sc, Zn, Ag, and Cd) and 8 (Sc, Cu-Zr, Ru, Ag, and Cd), respectively.
Regarding the influence of non-metal constituents, M$_2$A$_2$B altermagnets (Fig. 2a) show a clear preference for either halogens occupying site A or oxygens occupying site B.
Conversely, Janus M$_2$AA$'$B altermagnets (Fig. 2c) exhibit much weaker dependence on non-metal constituents, although a slightly higher prevalence is observed when halogens or pnictogens occupy the A/A' sites.
In contrast, the formation of altermagnetism in M$_2$A$_2$ (Fig. 2b) and M$_2$AA$'$ (Fig. 2d) appears to be largely insensitive to the A-site elements, with M = Ni, Zr, Rh, and Pd being exceptional cases that favor halogens at site A.
These element-dependent trends can be attributed to the interplay between exchange interactions governed by the Goodenough–Kanamori–Anderson rules \cite{Goodenough1958,Kanamori1959,Anderson1955}, and to the variations in $d$-shell electron filling arising from different chemical bonding environments and orbital hybridization.

The emergence of Dirac-cone band structures do have a strong dependence on the presence of AM order, as evidenced by Fig. 2 and 3.
Approximately 74.6\% of candidates having Dirac-cone band structures are found to be AM,
however, the dependence strength varies by design frameworks.
The direct V$_2$Se$_2$O-analogous framework M$_2$A$_2$B has the weakest dependence, with only 44.0\% of its Dirac-cone candidates being AM.
In contrast, the modified and Janus structured exhibit significantly enhanced dependence,
namely 85.0\% in M$_2$A$_2$, 76.5\% in M$_2$AA$'$B, and peaking at 92.3\% in M$_2$AA$'$ that adopts both B-site removal and Janus structuring.
Such strong dependence likely stems from shared crystal-symmetry requirements for both Dirac-cone formation and altermagnetism,
which are specifically the two symmetry-related sublattices composed of the same chemical species \cite{Wang2015}. 
This suggests that the coexistence of Dirac cones and altermagnetism may be a frequent occurrence in other monolayer AM systems.

\subsection{Results of Example Candidates}

\begin{figure*}
   \centering
   \includegraphics[width=0.9\textwidth]{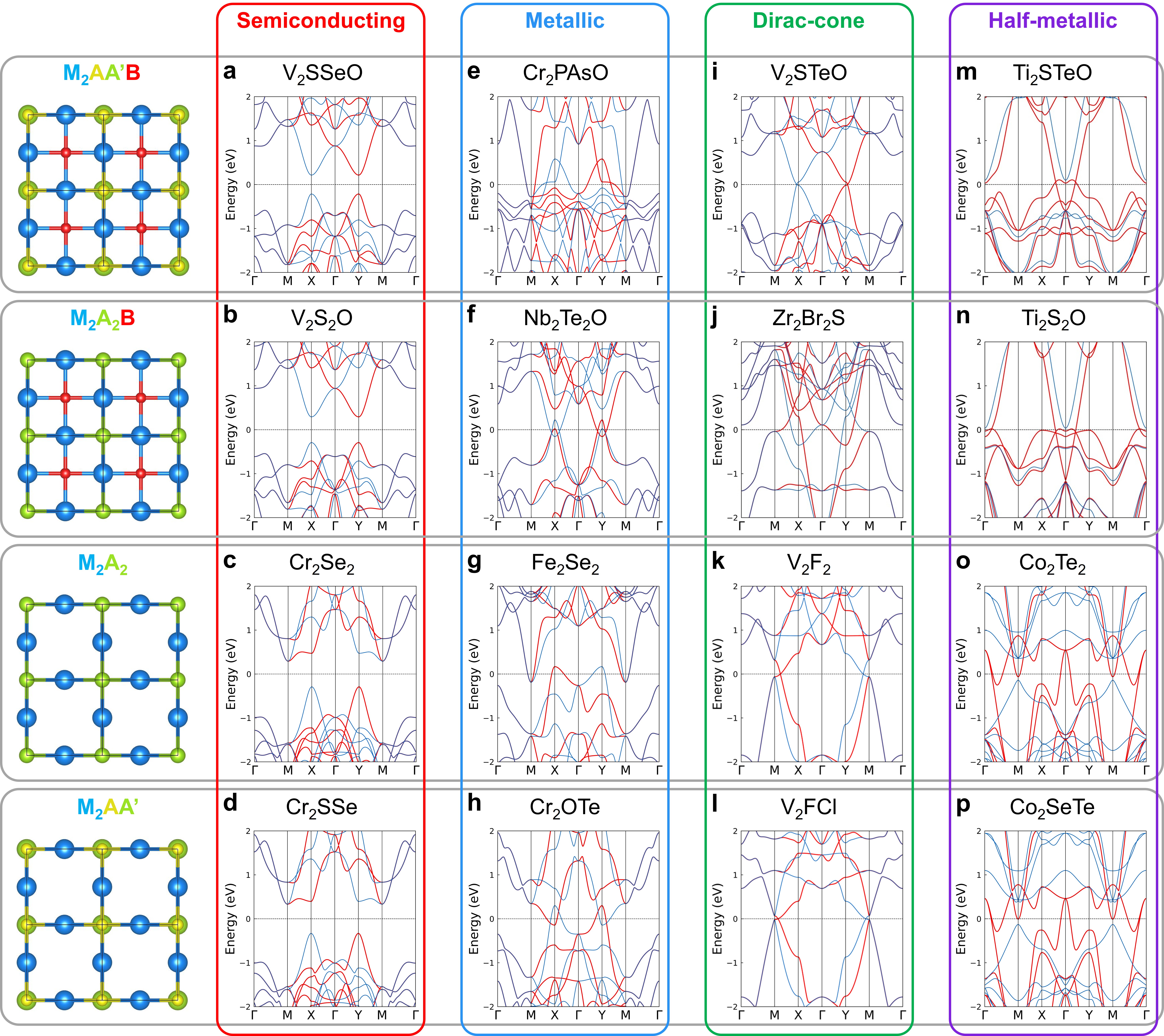}
   \caption{DFT-calculated spin-polarized band structures of selected AM candidates of material design frameworks (from top to bottom) Janus M$_2$AA$'$B, M$_2$A$_2$B, M$_2$A$_2$, and Janus M$_2$AA$'$, hosting (from left to right) semiconducting, metallic, Dirac-cone semimetallic, and half-metallic band features. The four types of band structures are marked at the top, while the top-view structures of four material frameworks are at the left, with the colored letters in chemical formula matching the respective atomic sites. AM ordering is the ground-state for all semiconducting, metallic, and Dirac-cone semimetallic candidates, while the FM ordering is for half-metallic ones. }
   \label{fig:4}
\end{figure*}
\begin{figure*}
   \centering
   \includegraphics[width=0.9\textwidth]{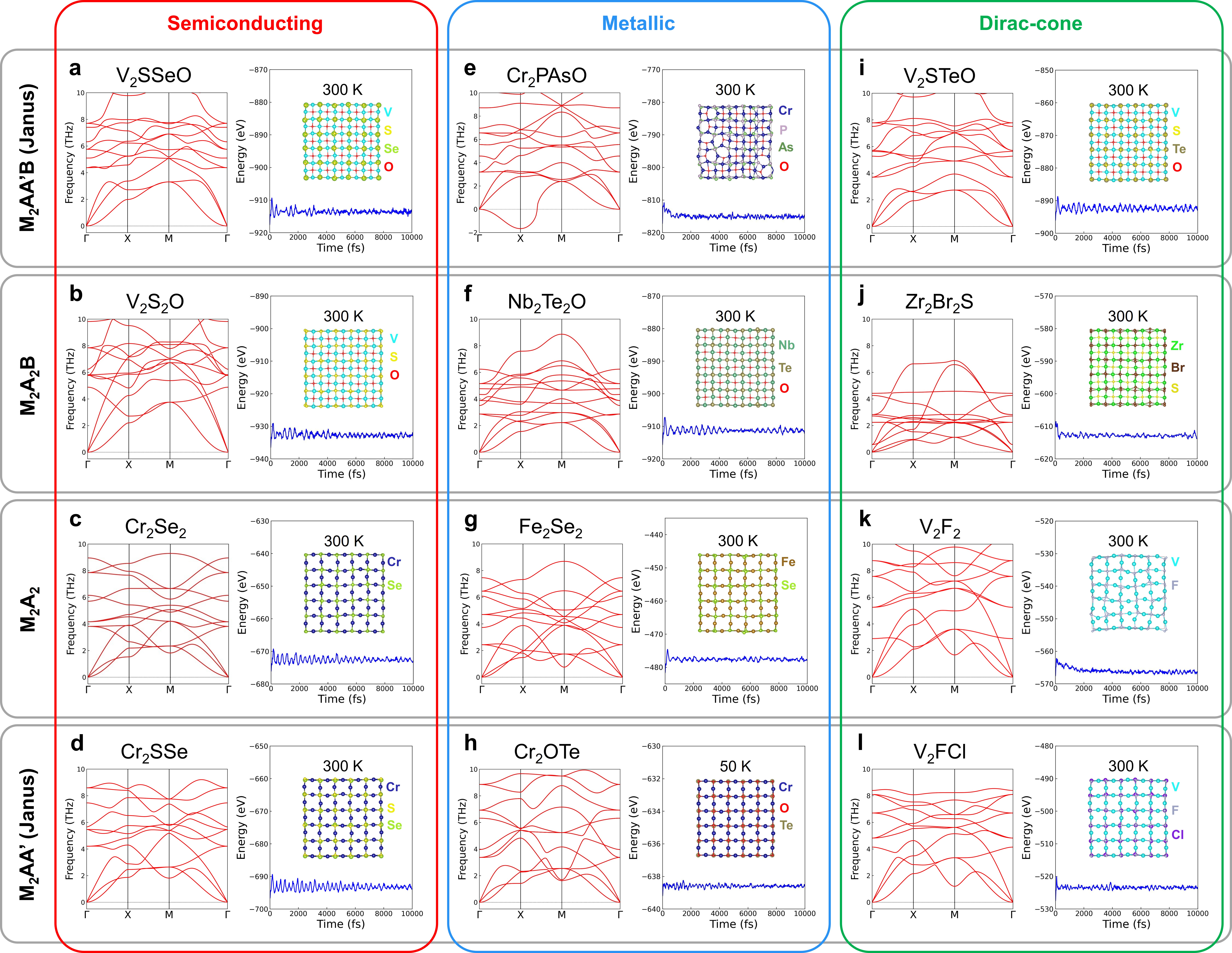}
   \caption{DFT-calculated phonon band structures (left) and AIMD simulations (right) for the same selected AM candidates of four material frameworks (from top to bottom) Janus M$_2$AA$'$B, M$_2$A$_2$B, M$_2$A$_2$, and Janus M$_2$AA$'$ with (from left to right) semiconducting, metallic, and Dirac-cone semimetallic band features. The top-view crystal structures after AIMD simulations at 300 K (except for 50 K in Cr$_2$OTe) after 10 ps are shown inset for different candidates, respectively, with colored letters on the right marking the names of ions. Almost all candidates presented exhibit both kinetic and thermodynamic stability. }
   \label{fig:5}
\end{figure*}

In Fig. 4, we selectively demonstrate examples of three primary spin-polarized electronic band structures under AM order,
namely, semiconducting, metallic, and Dirac-cone simimetallic,
using specific material candidates.
And the evolution of band structures also reflects the influence of element substitutions to electronic structures.
For semiconducting altermagnets (Fig. 4a-d),
element substitutions primarily alter the magnitude of band gaps while largely preserving the overall shapes of band dispersions.
For instance, in V$_2$S$_2$O,
substituting Se with less electronegative S at site A (compared to V$_2$Se$_2$O in Fig. 1e) weakens the V-S covalent bonding,
resulting in a narrowed band gap (Fig. 4b).
Conversely, substituting V with Cr at site M in Cr$_2$Se$_2$ (compared to V$_2$Se$_2$ in Fig. 1g) enhances covalent bonding,
as the $d$-shell filling of Cr$^{2+}$ ($3d^4$) is closer to half-filling that that of V$^{2+}$ ($3d^3$),
leading to an enlarged band gap (Fig. 4c).
Furthermore, the Janus structuring in V$_2$SSeO (Fig. 4a) and Cr$_2$SSe (Fig. 4d) only slightly modifies the gap magnitude due to the Stark effect arising from the Janus-induced structural asymmetry.

In contrast, for metallic altermagnets (Fig. 4e-h),
element substitutions significantly modify the bonding states and orbital hybridization,
causing shifts in both conduction and valence bands that lead them to intersect the Fermi level.
Specifically, in Nb$_2$Te$_2$O (Fig. 4f), the simultaneous substitutions at both site M and A by chemically similar but more electropositive elements (V to Nb and Se to Te compared to V$_2$Se$_2$O) introduce more delocalized electrons that leads to metallic states from intersecting valleys at X point.
Substituting Cr with Fe at site M in Fe$_2$Se$_2$ (compared to Cr$_2$Se$_2$) introduces electrons exceeding half filling of $d$ shell (Fe$^{2+}$ with $3d^6$), which weakens the covalent bonding and also closes the gap, as shown in Fig. 4g.
The Janus structuring in metallic Cr$_2$PAsO (Fig. 4e) not only has shifted conduction valley momentum due to V-to-Cr substitution,
but also gives rise to intersected, and more importantly,
strongly hybridized and reconstructed bands (slightly below the Fermi level near M point) due to Stark effect combined with less electronegative substitution by pnictogens.
In Janus Cr$_2$OTe (Fig. 4h), on the contrary, the Janus-induced Stark effect primarily shifts both conduction and valence band with larger energy to intersect with the Fermi level, compared with the case of Cr$_2$SSe.



The emergence of Dirac-cone band structures in altermagnets identified (Fig. 4i-l) involves not only substitution of more chemically active elements but also Stark effect-inducing Janus structuring with congeneric elements.
In both Zr$_2$Br$_2$S (Fig. 4j) and V$_2$F$_2$ (Fig. 4k),
the substitution of original constituent elements (V, Cr, Se, and O) with more chemically active ones results in significant interaction between conduction and valence states at/near the Fermi level,
leading to band-engineered formation of Dirac-cone linear dispersions (crossing points marked by black circles in Fig. S3-S17).
Differently, in V$_2$STeO (Fig. 4i),
the Dirac cones are induced primarily by Janus-induced Stark effect,
causing the spin-polarized valleys (originally at X/Y point of V$_2$Se$_2$O in Fig. 1e) to intersect precisely at a single point after shift.
The Dirac-cone band structures is preserved in V$_2$FCl after Janus structuring (Fig. 4l),
which is expected given that the chemical similarity between F and Cl,
as well as the generally negligible influence of the Stark effect on metallic or semimetallic states.

Nearly all altermagnets demonstrate in Fig. 4 exhibit stability from both kinetic and thermodynamic perspective,
as evidenced by first-principles phonon calculations and $ab$-initio molecular dynamics (AIMD) simulations.
As shown in Fig. 5, most candidates have no imaginary phonon modes and maintain atomic structures and energies at room temperature (about 300 K for over 10 ps),
with only two exceptions Cr$_2$PAsO and Cr$_2$OTe.
Cr$_2$PAsO (Fig. 5e) shows large imaginary phonon modes, indicating kinetic instability despite the seemingly stable AIMD results.
Cr$_2$OTe (Fig. 5h), while having no imaginary phonon modes, only demonstrates structural stability in AIMD simulations at a low temperature (50 K).
The confirmed stability of these altermagnets highlights their potential feasibility in future experimental synthesis and in-depth investigations.

In addition to band structures under AM order,
Fig. 4m-p also selectively present examples of spintronics-favored half-metallic ones under FM order.
The single-spin metallic character showcased by materials like Ti$_2$S$_2$O (Fig. 4n) and Co$_2$Te$_2$ (Fig. 4o) arises directly from band engineering.
This involves not only weakened covalent bonding resulting from substitution with certain chemically active elements,
but also altered $d$-shell filling due to the interplay between excess $d$ electrons and FM crystal-field splitting.
Notably, in contrast to the AM cases, subsequent Janus structuring exerts negligible influence on these half-metallic band structures (Fig. 4m and p).
Such negligible influence can be attributed to the relative ineffectiveness of intrinsic dipole field associated with the Stark effect in metallic systems.

\section{Conclusions}

In summary, this study establishes design principles for monolayer altermagnets utilizing symmetry-preserving structural engineering and valence-adaptive element substitution.
Starting from the solely synthesized layered altermagnet V$_2$(Se,Te)$_2$O as the structure template,
we derive four distinct structural frameworks: M$_2$A$_2$B, M$_2$A$_2$, M$_2$AA$'$B, and M$_2$AA$'$.
They are generated either by removing of site B or Janus structuring at site A while preserving the critical symmetry required by altermagnetism.
On this basis, we construct 2600 candidates in total by substituting elements at all atomic sites upon satisfying valence-adaptive and closed-shell condition.
The elements occupying site M, A/A$'$, and B are taken from the first 20 transition metals, the first 4 non-metal elements of group VA-VIIA, and the first 4 chalcogens, respectively.
Further high-throughput first-principles calculations determine their ground-state magnetic orderings and electronic band structures,
and identify 670 altermagnets representing approximately one quarter of the designed candidates.
Such highly predictive accuracy proves the rationality and effectiveness of our design principles in searching atomically thin altermagnets and implies potential generaslization to other similar systems.
Notably, 91 of these altermagnets simultaneously host CSML Dirac-cone band structures near the Fermi level, 
the coexistence of AM order and CSML Dirac cones gives rise to anisotropic and spin-selective ultra-fast carrier transport,
thus promising in spintronics research and applications.
Our designs and calculations not only provide plenty of highly realizable monolayer altermagnets and their diverse electronic properties to guide future investigations,
but also establishes generalizable principles for their rational design.

\section{Computational Methods}

The high-throughput first-principles calculations in the chemical design are performed by using the first-principles density functional theory (DFT) methods implemented in the Vienna $ab$-initio Simulation Package (VASP) \cite{KRESSE199615,PhysRevB.54.11169}.
The exchange correlations and the ion-electron interactions are treated by the Perdew-Burke-Ernzerhof (PBE) functional \cite{PhysRevLett.77.3865} and the projected-augmented wave (PAW) \cite{PhysRevB.50.17953,PhysRevB.59.1758} under the Generalized Gradient Approximation (GGA).
The vacuum space between the periodic images along the $z$ axis is no less than 12 \AA~to avoid unwanted interactions from neighboring van der Waals layers.
Throughout all calculations, the convergence criteria for energy and force are less than $1\times10^{-5}$ eV and $-0.01$ eV/\AA, and the plane-wave energy cutoff is kept as 600 eV.
The k-point sampling uses a Monkhorst-Pack scheme \cite{PhysRevB.93.155109} with an accuracy of $0.06\pi/$\AA~and $0.02\pi/$\AA~for structural relaxations and self-consistent electronic calculations, respectively.
To obtain reasonable electronic structures and magnetic orderings, the on-site Coulomb interaction is considered by adding effective Hubbard U on each transition-metal element following Ref. \cite{DFTU_1}.
The phonon band structures are calculated using the density functional perturbation theory (DFPT) \cite{baroni2001} implemented in VASP, with the help of the PHONOPY package \cite{phonopy-phono3py-JPCM,phonopy-phono3py-JPSJ}.
The pseudo-imaginary modes in few candidates may come from anharmonic higher-order phonons and are corrected by the Hiphive package \cite{PhysRevB.77.144112,PhysRevB.84.180301}.
The ab-initio molecular dynamics (AIMD) simulations are also performed by VASP using a Nose-Hoover thermostat \cite{Nos1984,Nos1991,Hoover1985} for no less than 10 ps.


\section{Supplementary Data}
Supplementary data are available online.

\section{Acknowledgments}
This work is supported by National Key R$\&$D Program of China (2021YFA1401500) and the Hong Kong Research Grants Council (16303821, 16306722, 16304523 and C6046-24G).



~

\textbf{\textit{Conflict of interest statement.}} None declared.

\bibliography{ref}

\begin{thebibliography}{71}%
\makeatletter
\providecommand \@ifxundefined [1]{%
 \@ifx{#1\undefined}
}%
\providecommand \@ifnum [1]{%
 \ifnum #1\expandafter \@firstoftwo
 \else \expandafter \@secondoftwo
 \fi
}%
\providecommand \@ifx [1]{%
 \ifx #1\expandafter \@firstoftwo
 \else \expandafter \@secondoftwo
 \fi
}%
\providecommand \natexlab [1]{#1}%
\providecommand \enquote  [1]{``#1''}%
\providecommand \bibnamefont  [1]{#1}%
\providecommand \bibfnamefont [1]{#1}%
\providecommand \citenamefont [1]{#1}%
\providecommand \href@noop [0]{\@secondoftwo}%
\providecommand \href [0]{\begingroup \@sanitize@url \@href}%
\providecommand \@href[1]{\@@startlink{#1}\@@href}%
\providecommand \@@href[1]{\endgroup#1\@@endlink}%
\providecommand \@sanitize@url [0]{\catcode `\\12\catcode `\$12\catcode
  `\&12\catcode `\#12\catcode `\^12\catcode `\_12\catcode `\%12\relax}%
\providecommand \@@startlink[1]{}%
\providecommand \@@endlink[0]{}%
\providecommand \url  [0]{\begingroup\@sanitize@url \@url }%
\providecommand \@url [1]{\endgroup\@href {#1}{\urlprefix }}%
\providecommand \urlprefix  [0]{URL }%
\providecommand \Eprint [0]{\href }%
\providecommand \doibase [0]{https://doi.org/}%
\providecommand \selectlanguage [0]{\@gobble}%
\providecommand \bibinfo  [0]{\@secondoftwo}%
\providecommand \bibfield  [0]{\@secondoftwo}%
\providecommand \translation [1]{[#1]}%
\providecommand \BibitemOpen [0]{}%
\providecommand \bibitemStop [0]{}%
\providecommand \bibitemNoStop [0]{.\EOS\space}%
\providecommand \EOS [0]{\spacefactor3000\relax}%
\providecommand \BibitemShut  [1]{\csname bibitem#1\endcsname}%
\let\auto@bib@innerbib\@empty
\bibitem [{\citenamefont {Wu}\ \emph {et~al.}(2007)\citenamefont {Wu},
  \citenamefont {Sun}, \citenamefont {Fradkin},\ and\ \citenamefont
  {Zhang}}]{CJWu2009}%
  \BibitemOpen
  \bibfield  {author} {\bibinfo {author} {\bibfnamefont {C.}~\bibnamefont
  {Wu}}, \bibinfo {author} {\bibfnamefont {K.}~\bibnamefont {Sun}}, \bibinfo
  {author} {\bibfnamefont {E.}~\bibnamefont {Fradkin}},\ and\ \bibinfo {author}
  {\bibfnamefont {S.-C.}\ \bibnamefont {Zhang}},\ }\bibfield  {title} {\bibinfo
  {title} {Fermi liquid instabilities in the spin channel},\ }\href
  {https://doi.org/10.1103/PhysRevB.75.115103} {\bibfield  {journal} {\bibinfo
  {journal} {Phys. Rev. B}\ }\textbf {\bibinfo {volume} {75}},\ \bibinfo
  {pages} {115103} (\bibinfo {year} {2007})}\BibitemShut {NoStop}%
\bibitem [{\citenamefont {Naka}\ \emph {et~al.}(2019)\citenamefont {Naka},
  \citenamefont {Hayami}, \citenamefont {Kusunose}, \citenamefont {Yanagi},
  \citenamefont {Motome},\ and\ \citenamefont {Seo}}]{naka2019spin}%
  \BibitemOpen
  \bibfield  {author} {\bibinfo {author} {\bibfnamefont {M.}~\bibnamefont
  {Naka}}, \bibinfo {author} {\bibfnamefont {S.}~\bibnamefont {Hayami}},
  \bibinfo {author} {\bibfnamefont {H.}~\bibnamefont {Kusunose}}, \bibinfo
  {author} {\bibfnamefont {Y.}~\bibnamefont {Yanagi}}, \bibinfo {author}
  {\bibfnamefont {Y.}~\bibnamefont {Motome}},\ and\ \bibinfo {author}
  {\bibfnamefont {H.}~\bibnamefont {Seo}},\ }\bibfield  {title} {\bibinfo
  {title} {Spin current generation in organic antiferromagnets},\ }\href
  {https://doi.org/10.1038/s41467-019-12229-y} {\bibfield  {journal} {\bibinfo
  {journal} {Nature communications}\ }\textbf {\bibinfo {volume} {10}},\
  \bibinfo {pages} {4305} (\bibinfo {year} {2019})}\BibitemShut {NoStop}%
\bibitem [{\citenamefont {Hayami}\ \emph {et~al.}(2019)\citenamefont {Hayami},
  \citenamefont {Yanagi},\ and\ \citenamefont {Kusunose}}]{Hayami2019}%
  \BibitemOpen
  \bibfield  {author} {\bibinfo {author} {\bibfnamefont {S.}~\bibnamefont
  {Hayami}}, \bibinfo {author} {\bibfnamefont {Y.}~\bibnamefont {Yanagi}},\
  and\ \bibinfo {author} {\bibfnamefont {H.}~\bibnamefont {Kusunose}},\
  }\bibfield  {title} {\bibinfo {title} {Momentum-dependent spin splitting by
  collinear antiferromagnetic ordering},\ }\href
  {https://doi.org/10.7566/jpsj.88.123702} {\bibfield  {journal} {\bibinfo
  {journal} {Journal of the Physical Society of Japan}\ }\textbf {\bibinfo
  {volume} {88}},\ \bibinfo {pages} {123702} (\bibinfo {year}
  {2019})}\BibitemShut {NoStop}%
\bibitem [{\citenamefont {Šmejkal}\ \emph {et~al.}(2020)\citenamefont
  {Šmejkal}, \citenamefont {González-Hernández}, \citenamefont {Jungwirth},\
  and\ \citenamefont {Sinova}}]{sciadv_altermag}%
  \BibitemOpen
  \bibfield  {author} {\bibinfo {author} {\bibfnamefont {L.}~\bibnamefont
  {Šmejkal}}, \bibinfo {author} {\bibfnamefont {R.}~\bibnamefont
  {González-Hernández}}, \bibinfo {author} {\bibfnamefont {T.}~\bibnamefont
  {Jungwirth}},\ and\ \bibinfo {author} {\bibfnamefont {J.}~\bibnamefont
  {Sinova}},\ }\bibfield  {title} {\bibinfo {title} {Crystal time-reversal
  symmetry breaking and spontaneous hall effect in collinear
  antiferromagnets},\ }\href {https://doi.org/10.1126/sciadv.aaz8809}
  {\bibfield  {journal} {\bibinfo  {journal} {Science Advances}\ }\textbf
  {\bibinfo {volume} {6}},\ \bibinfo {pages} {eaaz8809} (\bibinfo {year}
  {2020})}\BibitemShut {NoStop}%
\bibitem [{\citenamefont {Yuan}\ \emph {et~al.}(2021)\citenamefont {Yuan},
  \citenamefont {Wang}, \citenamefont {Luo},\ and\ \citenamefont
  {Zunger}}]{YLDong2019}%
  \BibitemOpen
  \bibfield  {author} {\bibinfo {author} {\bibfnamefont {L.-D.}\ \bibnamefont
  {Yuan}}, \bibinfo {author} {\bibfnamefont {Z.}~\bibnamefont {Wang}}, \bibinfo
  {author} {\bibfnamefont {J.-W.}\ \bibnamefont {Luo}},\ and\ \bibinfo {author}
  {\bibfnamefont {A.}~\bibnamefont {Zunger}},\ }\bibfield  {title} {\bibinfo
  {title} {Prediction of low-z collinear and noncollinear antiferromagnetic
  compounds having momentum-dependent spin splitting even without spin-orbit
  coupling},\ }\href {https://doi.org/10.1103/PhysRevMaterials.5.014409}
  {\bibfield  {journal} {\bibinfo  {journal} {Phys. Rev. Mater.}\ }\textbf
  {\bibinfo {volume} {5}},\ \bibinfo {pages} {014409} (\bibinfo {year}
  {2021})}\BibitemShut {NoStop}%
\bibitem [{\citenamefont {Ma}\ \emph {et~al.}(2021)\citenamefont {Ma},
  \citenamefont {Hu}, \citenamefont {Li}, \citenamefont {Liu}, \citenamefont
  {Yao}, \citenamefont {Jia},\ and\ \citenamefont
  {Liu}}]{ma2021multifunctional}%
  \BibitemOpen
  \bibfield  {author} {\bibinfo {author} {\bibfnamefont {H.-Y.}\ \bibnamefont
  {Ma}}, \bibinfo {author} {\bibfnamefont {M.}~\bibnamefont {Hu}}, \bibinfo
  {author} {\bibfnamefont {N.}~\bibnamefont {Li}}, \bibinfo {author}
  {\bibfnamefont {J.}~\bibnamefont {Liu}}, \bibinfo {author} {\bibfnamefont
  {W.}~\bibnamefont {Yao}}, \bibinfo {author} {\bibfnamefont {J.-F.}\
  \bibnamefont {Jia}},\ and\ \bibinfo {author} {\bibfnamefont {J.}~\bibnamefont
  {Liu}},\ }\bibfield  {title} {\bibinfo {title} {Multifunctional
  antiferromagnetic materials with giant piezomagnetism and noncollinear spin
  current},\ }\href {https://doi.org/10.1038/s41467-021-23127-7} {\bibfield
  {journal} {\bibinfo  {journal} {Nature communications}\ }\textbf {\bibinfo
  {volume} {12}},\ \bibinfo {pages} {2846} (\bibinfo {year}
  {2021})}\BibitemShut {NoStop}%
\bibitem [{\citenamefont {\ifmmode~\check{S}\else \v{S}\fi{}mejkal}\ \emph
  {et~al.}(2022{\natexlab{a}})\citenamefont {\ifmmode~\check{S}\else
  \v{S}\fi{}mejkal}, \citenamefont {Sinova},\ and\ \citenamefont
  {Jungwirth}}]{mejkal2022}%
  \BibitemOpen
  \bibfield  {author} {\bibinfo {author} {\bibfnamefont {L.}~\bibnamefont
  {\ifmmode~\check{S}\else \v{S}\fi{}mejkal}}, \bibinfo {author} {\bibfnamefont
  {J.}~\bibnamefont {Sinova}},\ and\ \bibinfo {author} {\bibfnamefont
  {T.}~\bibnamefont {Jungwirth}},\ }\bibfield  {title} {\bibinfo {title}
  {Beyond conventional ferromagnetism and antiferromagnetism: A phase with
  nonrelativistic spin and crystal rotation symmetry},\ }\href
  {https://doi.org/10.1103/PhysRevX.12.031042} {\bibfield  {journal} {\bibinfo
  {journal} {Phys. Rev. X}\ }\textbf {\bibinfo {volume} {12}},\ \bibinfo
  {pages} {031042} (\bibinfo {year} {2022}{\natexlab{a}})}\BibitemShut
  {NoStop}%
\bibitem [{\citenamefont {\ifmmode~\check{S}\else \v{S}\fi{}mejkal}\ \emph
  {et~al.}(2022{\natexlab{b}})\citenamefont {\ifmmode~\check{S}\else
  \v{S}\fi{}mejkal}, \citenamefont {Sinova},\ and\ \citenamefont
  {Jungwirth}}]{altermag_prx}%
  \BibitemOpen
  \bibfield  {author} {\bibinfo {author} {\bibfnamefont {L.}~\bibnamefont
  {\ifmmode~\check{S}\else \v{S}\fi{}mejkal}}, \bibinfo {author} {\bibfnamefont
  {J.}~\bibnamefont {Sinova}},\ and\ \bibinfo {author} {\bibfnamefont
  {T.}~\bibnamefont {Jungwirth}},\ }\bibfield  {title} {\bibinfo {title}
  {Emerging research landscape of altermagnetism},\ }\href
  {https://doi.org/10.1103/PhysRevX.12.040501} {\bibfield  {journal} {\bibinfo
  {journal} {Physical Review X}\ }\textbf {\bibinfo {volume} {12}},\ \bibinfo
  {pages} {040501} (\bibinfo {year} {2022}{\natexlab{b}})}\BibitemShut
  {NoStop}%
\bibitem [{\citenamefont {Song}\ \emph {et~al.}(2025)\citenamefont {Song},
  \citenamefont {Bai}, \citenamefont {Zhou}, \citenamefont {Han}, \citenamefont
  {Reichlova}, \citenamefont {Dil}, \citenamefont {Liu}, \citenamefont {Chen},\
  and\ \citenamefont {Pan}}]{song2025altermagnets}%
  \BibitemOpen
  \bibfield  {author} {\bibinfo {author} {\bibfnamefont {C.}~\bibnamefont
  {Song}}, \bibinfo {author} {\bibfnamefont {H.}~\bibnamefont {Bai}}, \bibinfo
  {author} {\bibfnamefont {Z.}~\bibnamefont {Zhou}}, \bibinfo {author}
  {\bibfnamefont {L.}~\bibnamefont {Han}}, \bibinfo {author} {\bibfnamefont
  {H.}~\bibnamefont {Reichlova}}, \bibinfo {author} {\bibfnamefont {J.~H.}\
  \bibnamefont {Dil}}, \bibinfo {author} {\bibfnamefont {J.}~\bibnamefont
  {Liu}}, \bibinfo {author} {\bibfnamefont {X.}~\bibnamefont {Chen}},\ and\
  \bibinfo {author} {\bibfnamefont {F.}~\bibnamefont {Pan}},\ }\bibfield
  {title} {\bibinfo {title} {Altermagnets as a new class of functional
  materials},\ }\bibfield  {journal} {\bibinfo  {journal} {Nature Reviews
  Materials}\ }\href {https://doi.org/10.1038/s41578-025-00779-1}
  {10.1038/s41578-025-00779-1} (\bibinfo {year} {2025})\BibitemShut {NoStop}%
\bibitem [{\citenamefont {Bai}\ \emph {et~al.}(2024)\citenamefont {Bai},
  \citenamefont {Feng}, \citenamefont {Liu}, \citenamefont {Šmejkal},
  \citenamefont {Mokrousov},\ and\ \citenamefont {Yao}}]{Bai2024}%
  \BibitemOpen
  \bibfield  {author} {\bibinfo {author} {\bibfnamefont {L.}~\bibnamefont
  {Bai}}, \bibinfo {author} {\bibfnamefont {W.}~\bibnamefont {Feng}}, \bibinfo
  {author} {\bibfnamefont {S.}~\bibnamefont {Liu}}, \bibinfo {author}
  {\bibfnamefont {L.}~\bibnamefont {Šmejkal}}, \bibinfo {author}
  {\bibfnamefont {Y.}~\bibnamefont {Mokrousov}},\ and\ \bibinfo {author}
  {\bibfnamefont {Y.}~\bibnamefont {Yao}},\ }\bibfield  {title} {\bibinfo
  {title} {Altermagnetism: Exploring new frontiers in magnetism and
  spintronics},\ }\href {https://doi.org/10.1002/adfm.202409327} {\bibfield
  {journal} {\bibinfo  {journal} {Advanced Functional Materials}\ }\textbf
  {\bibinfo {volume} {34}},\ \bibinfo {pages} {2409327} (\bibinfo {year}
  {2024})}\BibitemShut {NoStop}%
\bibitem [{\citenamefont {Fender}\ \emph {et~al.}(2025)\citenamefont {Fender},
  \citenamefont {Gonzalez},\ and\ \citenamefont {Bediako}}]{Fender2025}%
  \BibitemOpen
  \bibfield  {author} {\bibinfo {author} {\bibfnamefont {S.~S.}\ \bibnamefont
  {Fender}}, \bibinfo {author} {\bibfnamefont {O.}~\bibnamefont {Gonzalez}},\
  and\ \bibinfo {author} {\bibfnamefont {D.~K.}\ \bibnamefont {Bediako}},\
  }\bibfield  {title} {\bibinfo {title} {Altermagnetism: A chemical
  perspective},\ }\href {https://doi.org/10.1021/jacs.4c14503} {\bibfield
  {journal} {\bibinfo  {journal} {Journal of the American Chemical Society}\
  }\textbf {\bibinfo {volume} {147}},\ \bibinfo {pages} {2257–2274} (\bibinfo
  {year} {2025})}\BibitemShut {NoStop}%
\bibitem [{\citenamefont {Zhang}\ \emph {et~al.}(2025)\citenamefont {Zhang},
  \citenamefont {Cheng}, \citenamefont {Yin}, \citenamefont {Liu},
  \citenamefont {Deng}, \citenamefont {Qiao}, \citenamefont {Shi},
  \citenamefont {Zhang}, \citenamefont {Lin}, \citenamefont {Liu},
  \citenamefont {Ye}, \citenamefont {Huang}, \citenamefont {Meng},
  \citenamefont {Zhang}, \citenamefont {Okuda}, \citenamefont {Shimada},
  \citenamefont {Cui}, \citenamefont {Zhao}, \citenamefont {Cao}, \citenamefont
  {Qiao}, \citenamefont {Liu},\ and\ \citenamefont {Chen}}]{Zhang2025}%
  \BibitemOpen
  \bibfield  {author} {\bibinfo {author} {\bibfnamefont {F.}~\bibnamefont
  {Zhang}}, \bibinfo {author} {\bibfnamefont {X.}~\bibnamefont {Cheng}},
  \bibinfo {author} {\bibfnamefont {Z.}~\bibnamefont {Yin}}, \bibinfo {author}
  {\bibfnamefont {C.}~\bibnamefont {Liu}}, \bibinfo {author} {\bibfnamefont
  {L.}~\bibnamefont {Deng}}, \bibinfo {author} {\bibfnamefont {Y.}~\bibnamefont
  {Qiao}}, \bibinfo {author} {\bibfnamefont {Z.}~\bibnamefont {Shi}}, \bibinfo
  {author} {\bibfnamefont {S.}~\bibnamefont {Zhang}}, \bibinfo {author}
  {\bibfnamefont {J.}~\bibnamefont {Lin}}, \bibinfo {author} {\bibfnamefont
  {Z.}~\bibnamefont {Liu}}, \bibinfo {author} {\bibfnamefont {M.}~\bibnamefont
  {Ye}}, \bibinfo {author} {\bibfnamefont {Y.}~\bibnamefont {Huang}}, \bibinfo
  {author} {\bibfnamefont {X.}~\bibnamefont {Meng}}, \bibinfo {author}
  {\bibfnamefont {C.}~\bibnamefont {Zhang}}, \bibinfo {author} {\bibfnamefont
  {T.}~\bibnamefont {Okuda}}, \bibinfo {author} {\bibfnamefont
  {K.}~\bibnamefont {Shimada}}, \bibinfo {author} {\bibfnamefont
  {S.}~\bibnamefont {Cui}}, \bibinfo {author} {\bibfnamefont {Y.}~\bibnamefont
  {Zhao}}, \bibinfo {author} {\bibfnamefont {G.-H.}\ \bibnamefont {Cao}},
  \bibinfo {author} {\bibfnamefont {S.}~\bibnamefont {Qiao}}, \bibinfo {author}
  {\bibfnamefont {J.}~\bibnamefont {Liu}},\ and\ \bibinfo {author}
  {\bibfnamefont {C.}~\bibnamefont {Chen}},\ }\bibfield  {title} {\bibinfo
  {title} {Crystal-symmetry-paired spin-valley locking in a layered
  room-temperature metallic altermagnet candidate},\ }\bibfield  {journal}
  {\bibinfo  {journal} {Nature Physics}\ }\href
  {https://doi.org/10.1038/s41567-025-02864-2} {10.1038/s41567-025-02864-2}
  (\bibinfo {year} {2025})\BibitemShut {NoStop}%
\bibitem [{\citenamefont {Jiang}\ \emph {et~al.}(2025)\citenamefont {Jiang},
  \citenamefont {Hu}, \citenamefont {Bai}, \citenamefont {Song}, \citenamefont
  {Mu}, \citenamefont {Qu}, \citenamefont {Li}, \citenamefont {Zhu},
  \citenamefont {Pi}, \citenamefont {Wei}, \citenamefont {Sun}, \citenamefont
  {Huang}, \citenamefont {Zheng}, \citenamefont {Peng}, \citenamefont {He},
  \citenamefont {Li}, \citenamefont {Luo}, \citenamefont {Li}, \citenamefont
  {Chen}, \citenamefont {Li}, \citenamefont {Weng},\ and\ \citenamefont
  {Qian}}]{Jiang2025}%
  \BibitemOpen
  \bibfield  {author} {\bibinfo {author} {\bibfnamefont {B.}~\bibnamefont
  {Jiang}}, \bibinfo {author} {\bibfnamefont {M.}~\bibnamefont {Hu}}, \bibinfo
  {author} {\bibfnamefont {J.}~\bibnamefont {Bai}}, \bibinfo {author}
  {\bibfnamefont {Z.}~\bibnamefont {Song}}, \bibinfo {author} {\bibfnamefont
  {C.}~\bibnamefont {Mu}}, \bibinfo {author} {\bibfnamefont {G.}~\bibnamefont
  {Qu}}, \bibinfo {author} {\bibfnamefont {W.}~\bibnamefont {Li}}, \bibinfo
  {author} {\bibfnamefont {W.}~\bibnamefont {Zhu}}, \bibinfo {author}
  {\bibfnamefont {H.}~\bibnamefont {Pi}}, \bibinfo {author} {\bibfnamefont
  {Z.}~\bibnamefont {Wei}}, \bibinfo {author} {\bibfnamefont {Y.-J.}\
  \bibnamefont {Sun}}, \bibinfo {author} {\bibfnamefont {Y.}~\bibnamefont
  {Huang}}, \bibinfo {author} {\bibfnamefont {X.}~\bibnamefont {Zheng}},
  \bibinfo {author} {\bibfnamefont {Y.}~\bibnamefont {Peng}}, \bibinfo {author}
  {\bibfnamefont {L.}~\bibnamefont {He}}, \bibinfo {author} {\bibfnamefont
  {S.}~\bibnamefont {Li}}, \bibinfo {author} {\bibfnamefont {J.}~\bibnamefont
  {Luo}}, \bibinfo {author} {\bibfnamefont {Z.}~\bibnamefont {Li}}, \bibinfo
  {author} {\bibfnamefont {G.}~\bibnamefont {Chen}}, \bibinfo {author}
  {\bibfnamefont {H.}~\bibnamefont {Li}}, \bibinfo {author} {\bibfnamefont
  {H.}~\bibnamefont {Weng}},\ and\ \bibinfo {author} {\bibfnamefont
  {T.}~\bibnamefont {Qian}},\ }\bibfield  {title} {\bibinfo {title} {A metallic
  room-temperature d-wave altermagnet},\ }\bibfield  {journal} {\bibinfo
  {journal} {Nature Physics}\ }\href
  {https://doi.org/10.1038/s41567-025-02822-y} {10.1038/s41567-025-02822-y}
  (\bibinfo {year} {2025})\BibitemShut {NoStop}%
\bibitem [{\citenamefont {Tanaka}\ and\ \citenamefont
  {Golubov}(2007)}]{Tanaka2007}%
  \BibitemOpen
  \bibfield  {author} {\bibinfo {author} {\bibfnamefont {Y.}~\bibnamefont
  {Tanaka}}\ and\ \bibinfo {author} {\bibfnamefont {A.~A.}\ \bibnamefont
  {Golubov}},\ }\bibfield  {title} {\bibinfo {title} {Theory of the proximity
  effect in junctions with unconventional superconductors},\ }\href
  {https://doi.org/10.1103/physrevlett.98.037003} {\bibfield  {journal}
  {\bibinfo  {journal} {Physical Review Letters}\ }\textbf {\bibinfo {volume}
  {98}},\ \bibinfo {pages} {037003} (\bibinfo {year} {2007})}\BibitemShut
  {NoStop}%
\bibitem [{\citenamefont {Xu}\ \emph {et~al.}(2014)\citenamefont {Xu},
  \citenamefont {Liu}, \citenamefont {Wang}, \citenamefont {Ge}, \citenamefont
  {Liu}, \citenamefont {Yang}, \citenamefont {Chen}, \citenamefont {Liu},
  \citenamefont {Xu}, \citenamefont {Gao}, \citenamefont {Qian}, \citenamefont
  {Zhang},\ and\ \citenamefont {Jia}}]{Xu2014}%
  \BibitemOpen
  \bibfield  {author} {\bibinfo {author} {\bibfnamefont {J.-P.}\ \bibnamefont
  {Xu}}, \bibinfo {author} {\bibfnamefont {C.}~\bibnamefont {Liu}}, \bibinfo
  {author} {\bibfnamefont {M.-X.}\ \bibnamefont {Wang}}, \bibinfo {author}
  {\bibfnamefont {J.}~\bibnamefont {Ge}}, \bibinfo {author} {\bibfnamefont
  {Z.-L.}\ \bibnamefont {Liu}}, \bibinfo {author} {\bibfnamefont
  {X.}~\bibnamefont {Yang}}, \bibinfo {author} {\bibfnamefont {Y.}~\bibnamefont
  {Chen}}, \bibinfo {author} {\bibfnamefont {Y.}~\bibnamefont {Liu}}, \bibinfo
  {author} {\bibfnamefont {Z.-A.}\ \bibnamefont {Xu}}, \bibinfo {author}
  {\bibfnamefont {C.-L.}\ \bibnamefont {Gao}}, \bibinfo {author} {\bibfnamefont
  {D.}~\bibnamefont {Qian}}, \bibinfo {author} {\bibfnamefont {F.-C.}\
  \bibnamefont {Zhang}},\ and\ \bibinfo {author} {\bibfnamefont {J.-F.}\
  \bibnamefont {Jia}},\ }\bibfield  {title} {\bibinfo {title} {Artificial
  topological superconductor by the proximity effect},\ }\href
  {https://doi.org/10.1103/physrevlett.112.217001} {\bibfield  {journal}
  {\bibinfo  {journal} {Physical Review Letters}\ }\textbf {\bibinfo {volume}
  {112}},\ \bibinfo {pages} {217001} (\bibinfo {year} {2014})}\BibitemShut
  {NoStop}%
\bibitem [{\citenamefont {Wilson}\ \emph {et~al.}(1975)\citenamefont {Wilson},
  \citenamefont {Di~Salvo},\ and\ \citenamefont {Mahajan}}]{Wilson1975}%
  \BibitemOpen
  \bibfield  {author} {\bibinfo {author} {\bibfnamefont {J.}~\bibnamefont
  {Wilson}}, \bibinfo {author} {\bibfnamefont {F.}~\bibnamefont {Di~Salvo}},\
  and\ \bibinfo {author} {\bibfnamefont {S.}~\bibnamefont {Mahajan}},\
  }\bibfield  {title} {\bibinfo {title} {Charge-density waves and superlattices
  in the metallic layered transition metal dichalcogenides},\ }\href
  {https://doi.org/10.1080/00018737500101391} {\bibfield  {journal} {\bibinfo
  {journal} {Advances in Physics}\ }\textbf {\bibinfo {volume} {24}},\ \bibinfo
  {pages} {117–201} (\bibinfo {year} {1975})}\BibitemShut {NoStop}%
\bibitem [{\citenamefont {Mak}\ \emph {et~al.}(2010)\citenamefont {Mak},
  \citenamefont {Lee}, \citenamefont {Hone}, \citenamefont {Shan},\ and\
  \citenamefont {Heinz}}]{mak2010}%
  \BibitemOpen
  \bibfield  {author} {\bibinfo {author} {\bibfnamefont {K.~F.}\ \bibnamefont
  {Mak}}, \bibinfo {author} {\bibfnamefont {C.}~\bibnamefont {Lee}}, \bibinfo
  {author} {\bibfnamefont {J.}~\bibnamefont {Hone}}, \bibinfo {author}
  {\bibfnamefont {J.}~\bibnamefont {Shan}},\ and\ \bibinfo {author}
  {\bibfnamefont {T.~F.}\ \bibnamefont {Heinz}},\ }\bibfield  {title} {\bibinfo
  {title} {Atomically thin {M}o{S}2: A new direct-gap semiconductor},\ }\href
  {https://doi.org/10.1103/PhysRevLett.105.136805} {\bibfield  {journal}
  {\bibinfo  {journal} {Physical Review Letters}\ }\textbf {\bibinfo {volume}
  {105}},\ \bibinfo {pages} {136805} (\bibinfo {year} {2010})}\BibitemShut
  {NoStop}%
\bibitem [{\citenamefont {Qian}\ \emph {et~al.}(2014)\citenamefont {Qian},
  \citenamefont {Liu}, \citenamefont {Fu},\ and\ \citenamefont
  {Li}}]{Qian2014}%
  \BibitemOpen
  \bibfield  {author} {\bibinfo {author} {\bibfnamefont {X.}~\bibnamefont
  {Qian}}, \bibinfo {author} {\bibfnamefont {J.}~\bibnamefont {Liu}}, \bibinfo
  {author} {\bibfnamefont {L.}~\bibnamefont {Fu}},\ and\ \bibinfo {author}
  {\bibfnamefont {J.}~\bibnamefont {Li}},\ }\bibfield  {title} {\bibinfo
  {title} {Quantum spin hall effect in two-dimensional transition metal
  dichalcogenides},\ }\href {https://doi.org/10.1126/science.1256815}
  {\bibfield  {journal} {\bibinfo  {journal} {Science}\ }\textbf {\bibinfo
  {volume} {346}},\ \bibinfo {pages} {1344–1347} (\bibinfo {year}
  {2014})}\BibitemShut {NoStop}%
\bibitem [{\citenamefont {Mak}\ and\ \citenamefont {Shan}(2016)}]{Mak2016}%
  \BibitemOpen
  \bibfield  {author} {\bibinfo {author} {\bibfnamefont {K.~F.}\ \bibnamefont
  {Mak}}\ and\ \bibinfo {author} {\bibfnamefont {J.}~\bibnamefont {Shan}},\
  }\bibfield  {title} {\bibinfo {title} {Photonics and optoelectronics of 2{D}
  semiconductor transition metal dichalcogenides},\ }\href
  {https://doi.org/10.1038/nphoton.2015.282} {\bibfield  {journal} {\bibinfo
  {journal} {Nature Photonics}\ }\textbf {\bibinfo {volume} {10}},\ \bibinfo
  {pages} {216–226} (\bibinfo {year} {2016})}\BibitemShut {NoStop}%
\bibitem [{\citenamefont {Manzeli}\ \emph {et~al.}(2017)\citenamefont
  {Manzeli}, \citenamefont {Ovchinnikov}, \citenamefont {Pasquier},
  \citenamefont {Yazyev},\ and\ \citenamefont {Kis}}]{manzeli20172d}%
  \BibitemOpen
  \bibfield  {author} {\bibinfo {author} {\bibfnamefont {S.}~\bibnamefont
  {Manzeli}}, \bibinfo {author} {\bibfnamefont {D.}~\bibnamefont
  {Ovchinnikov}}, \bibinfo {author} {\bibfnamefont {D.}~\bibnamefont
  {Pasquier}}, \bibinfo {author} {\bibfnamefont {O.~V.}\ \bibnamefont
  {Yazyev}},\ and\ \bibinfo {author} {\bibfnamefont {A.}~\bibnamefont {Kis}},\
  }\bibfield  {title} {\bibinfo {title} {2{D} transition metal
  dichalcogenides},\ }\href {https://doi.org/10.1038/natrevmats.2017.33}
  {\bibfield  {journal} {\bibinfo  {journal} {Nature Reviews Materials}\
  }\textbf {\bibinfo {volume} {2}},\ \bibinfo {pages} {1} (\bibinfo {year}
  {2017})}\BibitemShut {NoStop}%
\bibitem [{\citenamefont {Li}\ \emph {et~al.}(2019)\citenamefont {Li},
  \citenamefont {Li}, \citenamefont {Du}, \citenamefont {Wang}, \citenamefont
  {Gu}, \citenamefont {Zhang}, \citenamefont {He}, \citenamefont {Duan},\ and\
  \citenamefont {Xu}}]{Li2019}%
  \BibitemOpen
  \bibfield  {author} {\bibinfo {author} {\bibfnamefont {J.}~\bibnamefont
  {Li}}, \bibinfo {author} {\bibfnamefont {Y.}~\bibnamefont {Li}}, \bibinfo
  {author} {\bibfnamefont {S.}~\bibnamefont {Du}}, \bibinfo {author}
  {\bibfnamefont {Z.}~\bibnamefont {Wang}}, \bibinfo {author} {\bibfnamefont
  {B.-L.}\ \bibnamefont {Gu}}, \bibinfo {author} {\bibfnamefont {S.-C.}\
  \bibnamefont {Zhang}}, \bibinfo {author} {\bibfnamefont {K.}~\bibnamefont
  {He}}, \bibinfo {author} {\bibfnamefont {W.}~\bibnamefont {Duan}},\ and\
  \bibinfo {author} {\bibfnamefont {Y.}~\bibnamefont {Xu}},\ }\bibfield
  {title} {\bibinfo {title} {Intrinsic magnetic topological insulators in van
  der waals layered {M}n{B}i2{T}e4-family materials},\ }\href
  {https://doi.org/10.1126/sciadv.aaw5685} {\bibfield  {journal} {\bibinfo
  {journal} {Science Advances}\ }\textbf {\bibinfo {volume} {5}},\ \bibinfo
  {pages} {eaaw5685} (\bibinfo {year} {2019})}\BibitemShut {NoStop}%
\bibitem [{\citenamefont {Gong}\ \emph {et~al.}(2019)\citenamefont {Gong},
  \citenamefont {Guo}, \citenamefont {Li}, \citenamefont {Zhu}, \citenamefont
  {Liao}, \citenamefont {Liu}, \citenamefont {Zhang}, \citenamefont {Gu},
  \citenamefont {Tang}, \citenamefont {Feng}, \citenamefont {Zhang},
  \citenamefont {Li}, \citenamefont {Song}, \citenamefont {Wang}, \citenamefont
  {Yu}, \citenamefont {Chen}, \citenamefont {Wang}, \citenamefont {Yao},
  \citenamefont {Duan}, \citenamefont {Xu}, \citenamefont {Zhang},
  \citenamefont {Ma}, \citenamefont {Xue},\ and\ \citenamefont
  {He}}]{Gong_2019}%
  \BibitemOpen
  \bibfield  {author} {\bibinfo {author} {\bibfnamefont {Y.}~\bibnamefont
  {Gong}}, \bibinfo {author} {\bibfnamefont {J.}~\bibnamefont {Guo}}, \bibinfo
  {author} {\bibfnamefont {J.}~\bibnamefont {Li}}, \bibinfo {author}
  {\bibfnamefont {K.}~\bibnamefont {Zhu}}, \bibinfo {author} {\bibfnamefont
  {M.}~\bibnamefont {Liao}}, \bibinfo {author} {\bibfnamefont {X.}~\bibnamefont
  {Liu}}, \bibinfo {author} {\bibfnamefont {Q.}~\bibnamefont {Zhang}}, \bibinfo
  {author} {\bibfnamefont {L.}~\bibnamefont {Gu}}, \bibinfo {author}
  {\bibfnamefont {L.}~\bibnamefont {Tang}}, \bibinfo {author} {\bibfnamefont
  {X.}~\bibnamefont {Feng}}, \bibinfo {author} {\bibfnamefont {D.}~\bibnamefont
  {Zhang}}, \bibinfo {author} {\bibfnamefont {W.}~\bibnamefont {Li}}, \bibinfo
  {author} {\bibfnamefont {C.}~\bibnamefont {Song}}, \bibinfo {author}
  {\bibfnamefont {L.}~\bibnamefont {Wang}}, \bibinfo {author} {\bibfnamefont
  {P.}~\bibnamefont {Yu}}, \bibinfo {author} {\bibfnamefont {X.}~\bibnamefont
  {Chen}}, \bibinfo {author} {\bibfnamefont {Y.}~\bibnamefont {Wang}}, \bibinfo
  {author} {\bibfnamefont {H.}~\bibnamefont {Yao}}, \bibinfo {author}
  {\bibfnamefont {W.}~\bibnamefont {Duan}}, \bibinfo {author} {\bibfnamefont
  {Y.}~\bibnamefont {Xu}}, \bibinfo {author} {\bibfnamefont {S.-C.}\
  \bibnamefont {Zhang}}, \bibinfo {author} {\bibfnamefont {X.}~\bibnamefont
  {Ma}}, \bibinfo {author} {\bibfnamefont {Q.-K.}\ \bibnamefont {Xue}},\ and\
  \bibinfo {author} {\bibfnamefont {K.}~\bibnamefont {He}},\ }\bibfield
  {title} {\bibinfo {title} {Experimental realization of an intrinsic magnetic
  topological insulator},\ }\href
  {https://doi.org/10.1088/0256-307X/36/7/076801} {\bibfield  {journal}
  {\bibinfo  {journal} {Chinese Physics Letters}\ }\textbf {\bibinfo {volume}
  {36}},\ \bibinfo {pages} {076801} (\bibinfo {year} {2019})}\BibitemShut
  {NoStop}%
\bibitem [{\citenamefont {Zhang}\ \emph {et~al.}(2019)\citenamefont {Zhang},
  \citenamefont {Shi}, \citenamefont {Zhu}, \citenamefont {Xing}, \citenamefont
  {Zhang},\ and\ \citenamefont {Wang}}]{Zhang2019}%
  \BibitemOpen
  \bibfield  {author} {\bibinfo {author} {\bibfnamefont {D.}~\bibnamefont
  {Zhang}}, \bibinfo {author} {\bibfnamefont {M.}~\bibnamefont {Shi}}, \bibinfo
  {author} {\bibfnamefont {T.}~\bibnamefont {Zhu}}, \bibinfo {author}
  {\bibfnamefont {D.}~\bibnamefont {Xing}}, \bibinfo {author} {\bibfnamefont
  {H.}~\bibnamefont {Zhang}},\ and\ \bibinfo {author} {\bibfnamefont
  {J.}~\bibnamefont {Wang}},\ }\bibfield  {title} {\bibinfo {title}
  {Topological axion states in the magnetic insulator {M}n{B}i2{T}e4 with the
  quantized magnetoelectric effect},\ }\href
  {https://doi.org/10.1103/physrevlett.122.206401} {\bibfield  {journal}
  {\bibinfo  {journal} {Physical Review Letters}\ }\textbf {\bibinfo {volume}
  {122}},\ \bibinfo {pages} {206401} (\bibinfo {year} {2019})}\BibitemShut
  {NoStop}%
\bibitem [{\citenamefont {He}\ and\ \citenamefont {Xue}(2019)}]{He2019}%
  \BibitemOpen
  \bibfield  {author} {\bibinfo {author} {\bibfnamefont {K.}~\bibnamefont
  {He}}\ and\ \bibinfo {author} {\bibfnamefont {Q.-K.}\ \bibnamefont {Xue}},\
  }\bibfield  {title} {\bibinfo {title} {The road to high-temperature quantum
  anomalous hall effect in magnetic topological insulators},\ }\href
  {https://doi.org/10.1142/s2010324719400162} {\bibfield  {journal} {\bibinfo
  {journal} {SPIN}\ }\textbf {\bibinfo {volume} {09}},\ \bibinfo {pages}
  {1940016} (\bibinfo {year} {2019})}\BibitemShut {NoStop}%
\bibitem [{\citenamefont {Deng}\ \emph {et~al.}(2020)\citenamefont {Deng},
  \citenamefont {Yu}, \citenamefont {Shi}, \citenamefont {Guo}, \citenamefont
  {Xu}, \citenamefont {Wang}, \citenamefont {Chen},\ and\ \citenamefont
  {Zhang}}]{Deng2020}%
  \BibitemOpen
  \bibfield  {author} {\bibinfo {author} {\bibfnamefont {Y.}~\bibnamefont
  {Deng}}, \bibinfo {author} {\bibfnamefont {Y.}~\bibnamefont {Yu}}, \bibinfo
  {author} {\bibfnamefont {M.~Z.}\ \bibnamefont {Shi}}, \bibinfo {author}
  {\bibfnamefont {Z.}~\bibnamefont {Guo}}, \bibinfo {author} {\bibfnamefont
  {Z.}~\bibnamefont {Xu}}, \bibinfo {author} {\bibfnamefont {J.}~\bibnamefont
  {Wang}}, \bibinfo {author} {\bibfnamefont {X.~H.}\ \bibnamefont {Chen}},\
  and\ \bibinfo {author} {\bibfnamefont {Y.}~\bibnamefont {Zhang}},\ }\bibfield
   {title} {\bibinfo {title} {Quantum anomalous hall effect in intrinsic
  magnetic topological insulator {M}n{B}i2{T}e4},\ }\href
  {https://doi.org/10.1126/science.aax8156} {\bibfield  {journal} {\bibinfo
  {journal} {Science}\ }\textbf {\bibinfo {volume} {367}},\ \bibinfo {pages}
  {895–900} (\bibinfo {year} {2020})}\BibitemShut {NoStop}%
\bibitem [{\citenamefont {Xiao}\ \emph {et~al.}(2012)\citenamefont {Xiao},
  \citenamefont {Liu}, \citenamefont {Feng}, \citenamefont {Xu},\ and\
  \citenamefont {Yao}}]{Xiao2012}%
  \BibitemOpen
  \bibfield  {author} {\bibinfo {author} {\bibfnamefont {D.}~\bibnamefont
  {Xiao}}, \bibinfo {author} {\bibfnamefont {G.-B.}\ \bibnamefont {Liu}},
  \bibinfo {author} {\bibfnamefont {W.}~\bibnamefont {Feng}}, \bibinfo {author}
  {\bibfnamefont {X.}~\bibnamefont {Xu}},\ and\ \bibinfo {author}
  {\bibfnamefont {W.}~\bibnamefont {Yao}},\ }\bibfield  {title} {\bibinfo
  {title} {Coupled spin and valley physics in monolayers of {M}o{S}2 and other
  group-{VI} dichalcogenides},\ }\href
  {https://doi.org/10.1103/physrevlett.108.196802} {\bibfield  {journal}
  {\bibinfo  {journal} {Physical Review Letters}\ }\textbf {\bibinfo {volume}
  {108}},\ \bibinfo {pages} {196802} (\bibinfo {year} {2012})}\BibitemShut
  {NoStop}%
\bibitem [{\citenamefont {Sipos}\ \emph {et~al.}(2008)\citenamefont {Sipos},
  \citenamefont {Kusmartseva}, \citenamefont {Akrap}, \citenamefont {Berger},
  \citenamefont {Forró},\ and\ \citenamefont {Tutiš}}]{Sipos2008}%
  \BibitemOpen
  \bibfield  {author} {\bibinfo {author} {\bibfnamefont {B.}~\bibnamefont
  {Sipos}}, \bibinfo {author} {\bibfnamefont {A.~F.}\ \bibnamefont
  {Kusmartseva}}, \bibinfo {author} {\bibfnamefont {A.}~\bibnamefont {Akrap}},
  \bibinfo {author} {\bibfnamefont {H.}~\bibnamefont {Berger}}, \bibinfo
  {author} {\bibfnamefont {L.}~\bibnamefont {Forró}},\ and\ \bibinfo {author}
  {\bibfnamefont {E.}~\bibnamefont {Tutiš}},\ }\bibfield  {title} {\bibinfo
  {title} {From {M}ott state to superconductivity in 1{T}-{T}a{S}2},\ }\href
  {https://doi.org/10.1038/nmat2318} {\bibfield  {journal} {\bibinfo  {journal}
  {Nature Materials}\ }\textbf {\bibinfo {volume} {7}},\ \bibinfo {pages}
  {960–965} (\bibinfo {year} {2008})}\BibitemShut {NoStop}%
\bibitem [{\citenamefont {Navarro-Moratalla}\ \emph {et~al.}(2016)\citenamefont
  {Navarro-Moratalla}, \citenamefont {Island}, \citenamefont {Mañas-Valero},
  \citenamefont {Pinilla-Cienfuegos}, \citenamefont {Castellanos-Gomez},
  \citenamefont {Quereda}, \citenamefont {Rubio-Bollinger}, \citenamefont
  {Chirolli}, \citenamefont {Silva-Guillén}, \citenamefont {Agraït},
  \citenamefont {Steele}, \citenamefont {Guinea}, \citenamefont {van~der
  Zant},\ and\ \citenamefont {Coronado}}]{NavarroMoratalla2016}%
  \BibitemOpen
  \bibfield  {author} {\bibinfo {author} {\bibfnamefont {E.}~\bibnamefont
  {Navarro-Moratalla}}, \bibinfo {author} {\bibfnamefont {J.~O.}\ \bibnamefont
  {Island}}, \bibinfo {author} {\bibfnamefont {S.}~\bibnamefont
  {Mañas-Valero}}, \bibinfo {author} {\bibfnamefont {E.}~\bibnamefont
  {Pinilla-Cienfuegos}}, \bibinfo {author} {\bibfnamefont {A.}~\bibnamefont
  {Castellanos-Gomez}}, \bibinfo {author} {\bibfnamefont {J.}~\bibnamefont
  {Quereda}}, \bibinfo {author} {\bibfnamefont {G.}~\bibnamefont
  {Rubio-Bollinger}}, \bibinfo {author} {\bibfnamefont {L.}~\bibnamefont
  {Chirolli}}, \bibinfo {author} {\bibfnamefont {J.~A.}\ \bibnamefont
  {Silva-Guillén}}, \bibinfo {author} {\bibfnamefont {N.}~\bibnamefont
  {Agraït}}, \bibinfo {author} {\bibfnamefont {G.~A.}\ \bibnamefont {Steele}},
  \bibinfo {author} {\bibfnamefont {F.}~\bibnamefont {Guinea}}, \bibinfo
  {author} {\bibfnamefont {H.~S.~J.}\ \bibnamefont {van~der Zant}},\ and\
  \bibinfo {author} {\bibfnamefont {E.}~\bibnamefont {Coronado}},\ }\bibfield
  {title} {\bibinfo {title} {Enhanced superconductivity in atomically thin
  {T}a{S}2},\ }\href {https://doi.org/10.1038/ncomms11043} {\bibfield
  {journal} {\bibinfo  {journal} {Nature Communications}\ }\textbf {\bibinfo
  {volume} {7}},\ \bibinfo {pages} {11043} (\bibinfo {year}
  {2016})}\BibitemShut {NoStop}%
\bibitem [{\citenamefont {Kusmartseva}\ \emph {et~al.}(2009)\citenamefont
  {Kusmartseva}, \citenamefont {Sipos}, \citenamefont {Berger}, \citenamefont
  {Forr\'o},\ and\ \citenamefont {Tuti\ifmmode~\check{s}\else
  \v{s}\fi{}}}]{Kusmartseva2009}%
  \BibitemOpen
  \bibfield  {author} {\bibinfo {author} {\bibfnamefont {A.~F.}\ \bibnamefont
  {Kusmartseva}}, \bibinfo {author} {\bibfnamefont {B.}~\bibnamefont {Sipos}},
  \bibinfo {author} {\bibfnamefont {H.}~\bibnamefont {Berger}}, \bibinfo
  {author} {\bibfnamefont {L.}~\bibnamefont {Forr\'o}},\ and\ \bibinfo {author}
  {\bibfnamefont {E.}~\bibnamefont {Tuti\ifmmode~\check{s}\else \v{s}\fi{}}},\
  }\bibfield  {title} {\bibinfo {title} {Pressure induced superconductivity in
  pristine 1{T}-{T}i{S}e2},\ }\href
  {https://doi.org/10.1103/PhysRevLett.103.236401} {\bibfield  {journal}
  {\bibinfo  {journal} {Physical Review Letters}\ }\textbf {\bibinfo {volume}
  {103}},\ \bibinfo {pages} {236401} (\bibinfo {year} {2009})}\BibitemShut
  {NoStop}%
\bibitem [{\citenamefont {Calandra}\ and\ \citenamefont
  {Mauri}(2011)}]{Calandra2011}%
  \BibitemOpen
  \bibfield  {author} {\bibinfo {author} {\bibfnamefont {M.}~\bibnamefont
  {Calandra}}\ and\ \bibinfo {author} {\bibfnamefont {F.}~\bibnamefont
  {Mauri}},\ }\bibfield  {title} {\bibinfo {title} {Charge-density wave and
  superconducting dome in {T}i{S}e2 from electron-phonon interaction},\ }\href
  {https://doi.org/10.1103/PhysRevLett.106.196406} {\bibfield  {journal}
  {\bibinfo  {journal} {Physical Review Letters}\ }\textbf {\bibinfo {volume}
  {106}},\ \bibinfo {pages} {196406} (\bibinfo {year} {2011})}\BibitemShut
  {NoStop}%
\bibitem [{\citenamefont {Joe}\ \emph {et~al.}(2014)\citenamefont {Joe},
  \citenamefont {Chen}, \citenamefont {Ghaemi}, \citenamefont {Finkelstein},
  \citenamefont {de~la Peña}, \citenamefont {Gan}, \citenamefont {Lee},
  \citenamefont {Yuan}, \citenamefont {Geck}, \citenamefont {MacDougall},
  \citenamefont {Chiang}, \citenamefont {Cooper}, \citenamefont {Fradkin},\
  and\ \citenamefont {Abbamonte}}]{Joe2014}%
  \BibitemOpen
  \bibfield  {author} {\bibinfo {author} {\bibfnamefont {Y.~I.}\ \bibnamefont
  {Joe}}, \bibinfo {author} {\bibfnamefont {X.~M.}\ \bibnamefont {Chen}},
  \bibinfo {author} {\bibfnamefont {P.}~\bibnamefont {Ghaemi}}, \bibinfo
  {author} {\bibfnamefont {K.~D.}\ \bibnamefont {Finkelstein}}, \bibinfo
  {author} {\bibfnamefont {G.~A.}\ \bibnamefont {de~la Peña}}, \bibinfo
  {author} {\bibfnamefont {Y.}~\bibnamefont {Gan}}, \bibinfo {author}
  {\bibfnamefont {J.~C.~T.}\ \bibnamefont {Lee}}, \bibinfo {author}
  {\bibfnamefont {S.}~\bibnamefont {Yuan}}, \bibinfo {author} {\bibfnamefont
  {J.}~\bibnamefont {Geck}}, \bibinfo {author} {\bibfnamefont {G.~J.}\
  \bibnamefont {MacDougall}}, \bibinfo {author} {\bibfnamefont {T.~C.}\
  \bibnamefont {Chiang}}, \bibinfo {author} {\bibfnamefont {S.~L.}\
  \bibnamefont {Cooper}}, \bibinfo {author} {\bibfnamefont {E.}~\bibnamefont
  {Fradkin}},\ and\ \bibinfo {author} {\bibfnamefont {P.}~\bibnamefont
  {Abbamonte}},\ }\bibfield  {title} {\bibinfo {title} {Emergence of charge
  density wave domain walls above the superconducting dome in 1{T}-{T}i{S}e2},\
  }\href {https://doi.org/10.1038/nphys2935} {\bibfield  {journal} {\bibinfo
  {journal} {Nature Physics}\ }\textbf {\bibinfo {volume} {10}},\ \bibinfo
  {pages} {421–425} (\bibinfo {year} {2014})}\BibitemShut {NoStop}%
\bibitem [{\citenamefont {Ali}\ \emph {et~al.}(2014)\citenamefont {Ali},
  \citenamefont {Xiong}, \citenamefont {Flynn}, \citenamefont {Tao},
  \citenamefont {Gibson}, \citenamefont {Schoop}, \citenamefont {Liang},
  \citenamefont {Haldolaarachchige}, \citenamefont {Hirschberger},
  \citenamefont {Ong},\ and\ \citenamefont {Cava}}]{Ali2014}%
  \BibitemOpen
  \bibfield  {author} {\bibinfo {author} {\bibfnamefont {M.~N.}\ \bibnamefont
  {Ali}}, \bibinfo {author} {\bibfnamefont {J.}~\bibnamefont {Xiong}}, \bibinfo
  {author} {\bibfnamefont {S.}~\bibnamefont {Flynn}}, \bibinfo {author}
  {\bibfnamefont {J.}~\bibnamefont {Tao}}, \bibinfo {author} {\bibfnamefont
  {Q.~D.}\ \bibnamefont {Gibson}}, \bibinfo {author} {\bibfnamefont {L.~M.}\
  \bibnamefont {Schoop}}, \bibinfo {author} {\bibfnamefont {T.}~\bibnamefont
  {Liang}}, \bibinfo {author} {\bibfnamefont {N.}~\bibnamefont
  {Haldolaarachchige}}, \bibinfo {author} {\bibfnamefont {M.}~\bibnamefont
  {Hirschberger}}, \bibinfo {author} {\bibfnamefont {N.~P.}\ \bibnamefont
  {Ong}},\ and\ \bibinfo {author} {\bibfnamefont {R.~J.}\ \bibnamefont
  {Cava}},\ }\bibfield  {title} {\bibinfo {title} {Large, non-saturating
  magnetoresistance in {WT}e2},\ }\href {https://doi.org/10.1038/nature13763}
  {\bibfield  {journal} {\bibinfo  {journal} {Nature}\ }\textbf {\bibinfo
  {volume} {514}},\ \bibinfo {pages} {205–208} (\bibinfo {year}
  {2014})}\BibitemShut {NoStop}%
\bibitem [{\citenamefont {Liu}\ \emph {et~al.}(2016{\natexlab{a}})\citenamefont
  {Liu}, \citenamefont {Wang}, \citenamefont {Fang}, \citenamefont {Fu},\ and\
  \citenamefont {Qian}}]{LiuJW2016}%
  \BibitemOpen
  \bibfield  {author} {\bibinfo {author} {\bibfnamefont {J.}~\bibnamefont
  {Liu}}, \bibinfo {author} {\bibfnamefont {H.}~\bibnamefont {Wang}}, \bibinfo
  {author} {\bibfnamefont {C.}~\bibnamefont {Fang}}, \bibinfo {author}
  {\bibfnamefont {L.}~\bibnamefont {Fu}},\ and\ \bibinfo {author}
  {\bibfnamefont {X.}~\bibnamefont {Qian}},\ }\bibfield  {title} {\bibinfo
  {title} {van der waals stacking-induced topological phase transition in
  layered ternary transition metal chalcogenides},\ }\href
  {https://doi.org/10.1021/acs.nanolett.6b04487} {\bibfield  {journal}
  {\bibinfo  {journal} {Nano Letters}\ }\textbf {\bibinfo {volume} {17}},\
  \bibinfo {pages} {467–475} (\bibinfo {year}
  {2016}{\natexlab{a}})}\BibitemShut {NoStop}%
\bibitem [{\citenamefont {Fei}\ \emph {et~al.}(2018)\citenamefont {Fei},
  \citenamefont {Zhao}, \citenamefont {Palomaki}, \citenamefont {Sun},
  \citenamefont {Miller}, \citenamefont {Zhao}, \citenamefont {Yan},
  \citenamefont {Xu},\ and\ \citenamefont {Cobden}}]{Fei2018}%
  \BibitemOpen
  \bibfield  {author} {\bibinfo {author} {\bibfnamefont {Z.}~\bibnamefont
  {Fei}}, \bibinfo {author} {\bibfnamefont {W.}~\bibnamefont {Zhao}}, \bibinfo
  {author} {\bibfnamefont {T.~A.}\ \bibnamefont {Palomaki}}, \bibinfo {author}
  {\bibfnamefont {B.}~\bibnamefont {Sun}}, \bibinfo {author} {\bibfnamefont
  {M.~K.}\ \bibnamefont {Miller}}, \bibinfo {author} {\bibfnamefont
  {Z.}~\bibnamefont {Zhao}}, \bibinfo {author} {\bibfnamefont {J.}~\bibnamefont
  {Yan}}, \bibinfo {author} {\bibfnamefont {X.}~\bibnamefont {Xu}},\ and\
  \bibinfo {author} {\bibfnamefont {D.~H.}\ \bibnamefont {Cobden}},\ }\bibfield
   {title} {\bibinfo {title} {Ferroelectric switching of a two-dimensional
  metal},\ }\href {https://doi.org/10.1038/s41586-018-0336-3} {\bibfield
  {journal} {\bibinfo  {journal} {Nature}\ }\textbf {\bibinfo {volume} {560}},\
  \bibinfo {pages} {336–339} (\bibinfo {year} {2018})}\BibitemShut {NoStop}%
\bibitem [{\citenamefont {Sharma}\ \emph {et~al.}(2019)\citenamefont {Sharma},
  \citenamefont {Xiang}, \citenamefont {Shao}, \citenamefont {Zhang},
  \citenamefont {Tsymbal}, \citenamefont {Hamilton},\ and\ \citenamefont
  {Seidel}}]{Sharma2019}%
  \BibitemOpen
  \bibfield  {author} {\bibinfo {author} {\bibfnamefont {P.}~\bibnamefont
  {Sharma}}, \bibinfo {author} {\bibfnamefont {F.-X.}\ \bibnamefont {Xiang}},
  \bibinfo {author} {\bibfnamefont {D.-F.}\ \bibnamefont {Shao}}, \bibinfo
  {author} {\bibfnamefont {D.}~\bibnamefont {Zhang}}, \bibinfo {author}
  {\bibfnamefont {E.~Y.}\ \bibnamefont {Tsymbal}}, \bibinfo {author}
  {\bibfnamefont {A.~R.}\ \bibnamefont {Hamilton}},\ and\ \bibinfo {author}
  {\bibfnamefont {J.}~\bibnamefont {Seidel}},\ }\bibfield  {title} {\bibinfo
  {title} {A room-temperature ferroelectric semimetal},\ }\href
  {https://doi.org/10.1126/sciadv.aax5080} {\bibfield  {journal} {\bibinfo
  {journal} {Science Advances}\ }\textbf {\bibinfo {volume} {5}},\ \bibinfo
  {pages} {eaax5080} (\bibinfo {year} {2019})}\BibitemShut {NoStop}%
\bibitem [{\citenamefont {Li}\ \emph {et~al.}(2017)\citenamefont {Li},
  \citenamefont {Wei}, \citenamefont {Zhao}, \citenamefont {Huang},\ and\
  \citenamefont {Dai}}]{Li2017}%
  \BibitemOpen
  \bibfield  {author} {\bibinfo {author} {\bibfnamefont {F.}~\bibnamefont
  {Li}}, \bibinfo {author} {\bibfnamefont {W.}~\bibnamefont {Wei}}, \bibinfo
  {author} {\bibfnamefont {P.}~\bibnamefont {Zhao}}, \bibinfo {author}
  {\bibfnamefont {B.}~\bibnamefont {Huang}},\ and\ \bibinfo {author}
  {\bibfnamefont {Y.}~\bibnamefont {Dai}},\ }\bibfield  {title} {\bibinfo
  {title} {Electronic and optical properties of pristine and vertical and
  lateral heterostructures of janus {M}o{SS}e and {WSS}e},\ }\href
  {https://doi.org/10.1021/acs.jpclett.7b02841} {\bibfield  {journal} {\bibinfo
   {journal} {The Journal of Physical Chemistry Letters}\ }\textbf {\bibinfo
  {volume} {8}},\ \bibinfo {pages} {5959–5965} (\bibinfo {year}
  {2017})}\BibitemShut {NoStop}%
\bibitem [{\citenamefont {Zhou}\ \emph {et~al.}(2019)\citenamefont {Zhou},
  \citenamefont {Chen}, \citenamefont {Yang}, \citenamefont {Liu},\ and\
  \citenamefont {Ouyang}}]{Zhou2019wsse}%
  \BibitemOpen
  \bibfield  {author} {\bibinfo {author} {\bibfnamefont {W.}~\bibnamefont
  {Zhou}}, \bibinfo {author} {\bibfnamefont {J.}~\bibnamefont {Chen}}, \bibinfo
  {author} {\bibfnamefont {Z.}~\bibnamefont {Yang}}, \bibinfo {author}
  {\bibfnamefont {J.}~\bibnamefont {Liu}},\ and\ \bibinfo {author}
  {\bibfnamefont {F.}~\bibnamefont {Ouyang}},\ }\bibfield  {title} {\bibinfo
  {title} {Geometry and electronic structure of monolayer, bilayer, and
  multilayer janus {WSS}e},\ }\href
  {https://doi.org/10.1103/PhysRevB.99.075160} {\bibfield  {journal} {\bibinfo
  {journal} {Physical Review B}\ }\textbf {\bibinfo {volume} {99}},\ \bibinfo
  {pages} {075160} (\bibinfo {year} {2019})}\BibitemShut {NoStop}%
\bibitem [{\citenamefont {Liu}\ \emph {et~al.}(2016{\natexlab{b}})\citenamefont
  {Liu}, \citenamefont {Zhang},\ and\ \citenamefont {Qi}}]{Liu2016}%
  \BibitemOpen
  \bibfield  {author} {\bibinfo {author} {\bibfnamefont {C.-X.}\ \bibnamefont
  {Liu}}, \bibinfo {author} {\bibfnamefont {S.-C.}\ \bibnamefont {Zhang}},\
  and\ \bibinfo {author} {\bibfnamefont {X.-L.}\ \bibnamefont {Qi}},\
  }\bibfield  {title} {\bibinfo {title} {The quantum anomalous hall effect:
  Theory and experiment},\ }\href
  {https://doi.org/10.1146/annurev-conmatphys-031115-011417} {\bibfield
  {journal} {\bibinfo  {journal} {Annual Review of Condensed Matter Physics}\
  }\textbf {\bibinfo {volume} {7}},\ \bibinfo {pages} {301–321} (\bibinfo
  {year} {2016}{\natexlab{b}})}\BibitemShut {NoStop}%
\bibitem [{\citenamefont {Chang}\ \emph {et~al.}(2023)\citenamefont {Chang},
  \citenamefont {Liu},\ and\ \citenamefont {MacDonald}}]{Chang2023}%
  \BibitemOpen
  \bibfield  {author} {\bibinfo {author} {\bibfnamefont {C.-Z.}\ \bibnamefont
  {Chang}}, \bibinfo {author} {\bibfnamefont {C.-X.}\ \bibnamefont {Liu}},\
  and\ \bibinfo {author} {\bibfnamefont {A.~H.}\ \bibnamefont {MacDonald}},\
  }\bibfield  {title} {\bibinfo {title} {Colloquium: Quantum anomalous hall
  effect},\ }\href {https://doi.org/10.1103/revmodphys.95.011002} {\bibfield
  {journal} {\bibinfo  {journal} {Reviews of Modern Physics}\ }\textbf
  {\bibinfo {volume} {95}},\ \bibinfo {pages} {011002} (\bibinfo {year}
  {2023})}\BibitemShut {NoStop}%
\bibitem [{\citenamefont {Yu}\ \emph {et~al.}(2010)\citenamefont {Yu},
  \citenamefont {Zhang}, \citenamefont {Zhang}, \citenamefont {Zhang},
  \citenamefont {Dai},\ and\ \citenamefont {Fang}}]{Yu2010}%
  \BibitemOpen
  \bibfield  {author} {\bibinfo {author} {\bibfnamefont {R.}~\bibnamefont
  {Yu}}, \bibinfo {author} {\bibfnamefont {W.}~\bibnamefont {Zhang}}, \bibinfo
  {author} {\bibfnamefont {H.-J.}\ \bibnamefont {Zhang}}, \bibinfo {author}
  {\bibfnamefont {S.-C.}\ \bibnamefont {Zhang}}, \bibinfo {author}
  {\bibfnamefont {X.}~\bibnamefont {Dai}},\ and\ \bibinfo {author}
  {\bibfnamefont {Z.}~\bibnamefont {Fang}},\ }\bibfield  {title} {\bibinfo
  {title} {Quantized anomalous hall effect in magnetic topological
  insulators},\ }\href {https://doi.org/10.1126/science.1187485} {\bibfield
  {journal} {\bibinfo  {journal} {Science}\ }\textbf {\bibinfo {volume}
  {329}},\ \bibinfo {pages} {61–64} (\bibinfo {year} {2010})}\BibitemShut
  {NoStop}%
\bibitem [{\citenamefont {Chang}\ \emph {et~al.}(2013)\citenamefont {Chang},
  \citenamefont {Zhang}, \citenamefont {Feng}, \citenamefont {Shen},
  \citenamefont {Zhang}, \citenamefont {Guo}, \citenamefont {Li}, \citenamefont
  {Ou}, \citenamefont {Wei}, \citenamefont {Wang}, \citenamefont {Ji},
  \citenamefont {Feng}, \citenamefont {Ji}, \citenamefont {Chen}, \citenamefont
  {Jia}, \citenamefont {Dai}, \citenamefont {Fang}, \citenamefont {Zhang},
  \citenamefont {He}, \citenamefont {Wang}, \citenamefont {Lu}, \citenamefont
  {Ma},\ and\ \citenamefont {Xue}}]{Chang2013}%
  \BibitemOpen
  \bibfield  {author} {\bibinfo {author} {\bibfnamefont {C.-Z.}\ \bibnamefont
  {Chang}}, \bibinfo {author} {\bibfnamefont {J.}~\bibnamefont {Zhang}},
  \bibinfo {author} {\bibfnamefont {X.}~\bibnamefont {Feng}}, \bibinfo {author}
  {\bibfnamefont {J.}~\bibnamefont {Shen}}, \bibinfo {author} {\bibfnamefont
  {Z.}~\bibnamefont {Zhang}}, \bibinfo {author} {\bibfnamefont
  {M.}~\bibnamefont {Guo}}, \bibinfo {author} {\bibfnamefont {K.}~\bibnamefont
  {Li}}, \bibinfo {author} {\bibfnamefont {Y.}~\bibnamefont {Ou}}, \bibinfo
  {author} {\bibfnamefont {P.}~\bibnamefont {Wei}}, \bibinfo {author}
  {\bibfnamefont {L.-L.}\ \bibnamefont {Wang}}, \bibinfo {author}
  {\bibfnamefont {Z.-Q.}\ \bibnamefont {Ji}}, \bibinfo {author} {\bibfnamefont
  {Y.}~\bibnamefont {Feng}}, \bibinfo {author} {\bibfnamefont {S.}~\bibnamefont
  {Ji}}, \bibinfo {author} {\bibfnamefont {X.}~\bibnamefont {Chen}}, \bibinfo
  {author} {\bibfnamefont {J.}~\bibnamefont {Jia}}, \bibinfo {author}
  {\bibfnamefont {X.}~\bibnamefont {Dai}}, \bibinfo {author} {\bibfnamefont
  {Z.}~\bibnamefont {Fang}}, \bibinfo {author} {\bibfnamefont {S.-C.}\
  \bibnamefont {Zhang}}, \bibinfo {author} {\bibfnamefont {K.}~\bibnamefont
  {He}}, \bibinfo {author} {\bibfnamefont {Y.}~\bibnamefont {Wang}}, \bibinfo
  {author} {\bibfnamefont {L.}~\bibnamefont {Lu}}, \bibinfo {author}
  {\bibfnamefont {X.-C.}\ \bibnamefont {Ma}},\ and\ \bibinfo {author}
  {\bibfnamefont {Q.-K.}\ \bibnamefont {Xue}},\ }\bibfield  {title} {\bibinfo
  {title} {Experimental observation of the quantum anomalous hall effect in a
  magnetic topological insulator},\ }\href
  {https://doi.org/10.1126/science.1234414} {\bibfield  {journal} {\bibinfo
  {journal} {Science}\ }\textbf {\bibinfo {volume} {340}},\ \bibinfo {pages}
  {167–170} (\bibinfo {year} {2013})}\BibitemShut {NoStop}%
\bibitem [{\citenamefont {Hsieh}\ \emph {et~al.}(2012)\citenamefont {Hsieh},
  \citenamefont {Lin}, \citenamefont {Liu}, \citenamefont {Duan}, \citenamefont
  {Bansil},\ and\ \citenamefont {Fu}}]{Hsieh2012}%
  \BibitemOpen
  \bibfield  {author} {\bibinfo {author} {\bibfnamefont {T.~H.}\ \bibnamefont
  {Hsieh}}, \bibinfo {author} {\bibfnamefont {H.}~\bibnamefont {Lin}}, \bibinfo
  {author} {\bibfnamefont {J.}~\bibnamefont {Liu}}, \bibinfo {author}
  {\bibfnamefont {W.}~\bibnamefont {Duan}}, \bibinfo {author} {\bibfnamefont
  {A.}~\bibnamefont {Bansil}},\ and\ \bibinfo {author} {\bibfnamefont
  {L.}~\bibnamefont {Fu}},\ }\bibfield  {title} {\bibinfo {title} {Topological
  crystalline insulators in the {S}n{T}e material class},\ }\href
  {https://doi.org/10.1038/ncomms1969} {\bibfield  {journal} {\bibinfo
  {journal} {Nature Communications}\ }\textbf {\bibinfo {volume} {3}},\
  \bibinfo {pages} {982} (\bibinfo {year} {2012})}\BibitemShut {NoStop}%
\bibitem [{\citenamefont {Tanaka}\ \emph {et~al.}(2012)\citenamefont {Tanaka},
  \citenamefont {Ren}, \citenamefont {Sato}, \citenamefont {Nakayama},
  \citenamefont {Souma}, \citenamefont {Takahashi}, \citenamefont {Segawa},\
  and\ \citenamefont {Ando}}]{Tanaka2012}%
  \BibitemOpen
  \bibfield  {author} {\bibinfo {author} {\bibfnamefont {Y.}~\bibnamefont
  {Tanaka}}, \bibinfo {author} {\bibfnamefont {Z.}~\bibnamefont {Ren}},
  \bibinfo {author} {\bibfnamefont {T.}~\bibnamefont {Sato}}, \bibinfo {author}
  {\bibfnamefont {K.}~\bibnamefont {Nakayama}}, \bibinfo {author}
  {\bibfnamefont {S.}~\bibnamefont {Souma}}, \bibinfo {author} {\bibfnamefont
  {T.}~\bibnamefont {Takahashi}}, \bibinfo {author} {\bibfnamefont
  {K.}~\bibnamefont {Segawa}},\ and\ \bibinfo {author} {\bibfnamefont
  {Y.}~\bibnamefont {Ando}},\ }\bibfield  {title} {\bibinfo {title}
  {Experimental realization of a topological crystalline insulator in
  {S}n{T}e},\ }\href {https://doi.org/10.1038/nphys2442} {\bibfield  {journal}
  {\bibinfo  {journal} {Nature Physics}\ }\textbf {\bibinfo {volume} {8}},\
  \bibinfo {pages} {800–803} (\bibinfo {year} {2012})}\BibitemShut {NoStop}%
\bibitem [{\citenamefont {Chang}\ \emph {et~al.}(2016)\citenamefont {Chang},
  \citenamefont {Liu}, \citenamefont {Lin}, \citenamefont {Wang}, \citenamefont
  {Zhao}, \citenamefont {Zhang}, \citenamefont {Jin}, \citenamefont {Zhong},
  \citenamefont {Hu}, \citenamefont {Duan}, \citenamefont {Zhang},
  \citenamefont {Fu}, \citenamefont {Xue}, \citenamefont {Chen},\ and\
  \citenamefont {Ji}}]{Chang2016}%
  \BibitemOpen
  \bibfield  {author} {\bibinfo {author} {\bibfnamefont {K.}~\bibnamefont
  {Chang}}, \bibinfo {author} {\bibfnamefont {J.}~\bibnamefont {Liu}}, \bibinfo
  {author} {\bibfnamefont {H.}~\bibnamefont {Lin}}, \bibinfo {author}
  {\bibfnamefont {N.}~\bibnamefont {Wang}}, \bibinfo {author} {\bibfnamefont
  {K.}~\bibnamefont {Zhao}}, \bibinfo {author} {\bibfnamefont {A.}~\bibnamefont
  {Zhang}}, \bibinfo {author} {\bibfnamefont {F.}~\bibnamefont {Jin}}, \bibinfo
  {author} {\bibfnamefont {Y.}~\bibnamefont {Zhong}}, \bibinfo {author}
  {\bibfnamefont {X.}~\bibnamefont {Hu}}, \bibinfo {author} {\bibfnamefont
  {W.}~\bibnamefont {Duan}}, \bibinfo {author} {\bibfnamefont {Q.}~\bibnamefont
  {Zhang}}, \bibinfo {author} {\bibfnamefont {L.}~\bibnamefont {Fu}}, \bibinfo
  {author} {\bibfnamefont {Q.-K.}\ \bibnamefont {Xue}}, \bibinfo {author}
  {\bibfnamefont {X.}~\bibnamefont {Chen}},\ and\ \bibinfo {author}
  {\bibfnamefont {S.-H.}\ \bibnamefont {Ji}},\ }\bibfield  {title} {\bibinfo
  {title} {Discovery of robust in-plane ferroelectricity in atomic-thick
  {S}n{T}e},\ }\href {https://doi.org/10.1126/science.aad8609} {\bibfield
  {journal} {\bibinfo  {journal} {Science}\ }\textbf {\bibinfo {volume}
  {353}},\ \bibinfo {pages} {274} (\bibinfo {year} {2016})}\BibitemShut
  {NoStop}%
\bibitem [{\citenamefont {Huang}\ \emph {et~al.}(2019)\citenamefont {Huang},
  \citenamefont {Nakamura}, \citenamefont {K\"{u}ster}, \citenamefont
  {Yaresko}, \citenamefont {Samal}, \citenamefont {Schr\"{o}ter}, \citenamefont
  {Strocov}, \citenamefont {Starke},\ and\ \citenamefont {Takagi}}]{Huang2019}%
  \BibitemOpen
  \bibfield  {author} {\bibinfo {author} {\bibfnamefont {D.}~\bibnamefont
  {Huang}}, \bibinfo {author} {\bibfnamefont {H.}~\bibnamefont {Nakamura}},
  \bibinfo {author} {\bibfnamefont {K.}~\bibnamefont {K\"{u}ster}}, \bibinfo
  {author} {\bibfnamefont {A.}~\bibnamefont {Yaresko}}, \bibinfo {author}
  {\bibfnamefont {D.}~\bibnamefont {Samal}}, \bibinfo {author} {\bibfnamefont
  {N.~B.~M.}\ \bibnamefont {Schr\"{o}ter}}, \bibinfo {author} {\bibfnamefont
  {V.~N.}\ \bibnamefont {Strocov}}, \bibinfo {author} {\bibfnamefont
  {U.}~\bibnamefont {Starke}},\ and\ \bibinfo {author} {\bibfnamefont
  {H.}~\bibnamefont {Takagi}},\ }\bibfield  {title} {\bibinfo {title} {Unusual
  valence state in the antiperovskites {S}r3{S}n{O} and {S}r3{P}b{O} revealed
  by {X}-ray photoelectron spectroscopy},\ }\href
  {https://doi.org/10.1103/physrevmaterials.3.124203} {\bibfield  {journal}
  {\bibinfo  {journal} {Physical Review Materials}\ }\textbf {\bibinfo {volume}
  {3}},\ \bibinfo {pages} {124203} (\bibinfo {year} {2019})}\BibitemShut
  {NoStop}%
\bibitem [{\citenamefont {Lam}\ \emph {et~al.}(2023)\citenamefont {Lam},
  \citenamefont {Peng}, \citenamefont {Wang}, \citenamefont {Wan},\ and\
  \citenamefont {Liu}}]{Lam2023}%
  \BibitemOpen
  \bibfield  {author} {\bibinfo {author} {\bibfnamefont {H.}~\bibnamefont
  {Lam}}, \bibinfo {author} {\bibfnamefont {R.}~\bibnamefont {Peng}}, \bibinfo
  {author} {\bibfnamefont {Z.}~\bibnamefont {Wang}}, \bibinfo {author}
  {\bibfnamefont {C.}~\bibnamefont {Wan}},\ and\ \bibinfo {author}
  {\bibfnamefont {J.}~\bibnamefont {Liu}},\ }\bibfield  {title} {\bibinfo
  {title} {Topological states of {S}r3{P}b{O}: From topological crystalline
  insulator phase in the bulk to quantum spin hall insulator phase in the
  thin-film limit},\ }\href {https://doi.org/10.1103/physrevb.108.155139}
  {\bibfield  {journal} {\bibinfo  {journal} {Physical Review B}\ }\textbf
  {\bibinfo {volume} {108}},\ \bibinfo {pages} {155139} (\bibinfo {year}
  {2023})}\BibitemShut {NoStop}%
\bibitem [{\citenamefont {Tang}\ \emph {et~al.}(2017)\citenamefont {Tang},
  \citenamefont {Zhang}, \citenamefont {Wong}, \citenamefont {Pedramrazi},
  \citenamefont {Tsai}, \citenamefont {Jia}, \citenamefont {Moritz},
  \citenamefont {Claassen}, \citenamefont {Ryu}, \citenamefont {Kahn},
  \citenamefont {Jiang}, \citenamefont {Yan}, \citenamefont {Hashimoto},
  \citenamefont {Lu}, \citenamefont {Moore}, \citenamefont {Hwang},
  \citenamefont {Hwang}, \citenamefont {Hussain}, \citenamefont {Chen},
  \citenamefont {Ugeda}, \citenamefont {Liu}, \citenamefont {Xie},
  \citenamefont {Devereaux}, \citenamefont {Crommie}, \citenamefont {Mo},\ and\
  \citenamefont {Shen}}]{Tang2017}%
  \BibitemOpen
  \bibfield  {author} {\bibinfo {author} {\bibfnamefont {S.}~\bibnamefont
  {Tang}}, \bibinfo {author} {\bibfnamefont {C.}~\bibnamefont {Zhang}},
  \bibinfo {author} {\bibfnamefont {D.}~\bibnamefont {Wong}}, \bibinfo {author}
  {\bibfnamefont {Z.}~\bibnamefont {Pedramrazi}}, \bibinfo {author}
  {\bibfnamefont {H.-Z.}\ \bibnamefont {Tsai}}, \bibinfo {author}
  {\bibfnamefont {C.}~\bibnamefont {Jia}}, \bibinfo {author} {\bibfnamefont
  {B.}~\bibnamefont {Moritz}}, \bibinfo {author} {\bibfnamefont
  {M.}~\bibnamefont {Claassen}}, \bibinfo {author} {\bibfnamefont
  {H.}~\bibnamefont {Ryu}}, \bibinfo {author} {\bibfnamefont {S.}~\bibnamefont
  {Kahn}}, \bibinfo {author} {\bibfnamefont {J.}~\bibnamefont {Jiang}},
  \bibinfo {author} {\bibfnamefont {H.}~\bibnamefont {Yan}}, \bibinfo {author}
  {\bibfnamefont {M.}~\bibnamefont {Hashimoto}}, \bibinfo {author}
  {\bibfnamefont {D.}~\bibnamefont {Lu}}, \bibinfo {author} {\bibfnamefont
  {R.~G.}\ \bibnamefont {Moore}}, \bibinfo {author} {\bibfnamefont {C.-C.}\
  \bibnamefont {Hwang}}, \bibinfo {author} {\bibfnamefont {C.}~\bibnamefont
  {Hwang}}, \bibinfo {author} {\bibfnamefont {Z.}~\bibnamefont {Hussain}},
  \bibinfo {author} {\bibfnamefont {Y.}~\bibnamefont {Chen}}, \bibinfo {author}
  {\bibfnamefont {M.~M.}\ \bibnamefont {Ugeda}}, \bibinfo {author}
  {\bibfnamefont {Z.}~\bibnamefont {Liu}}, \bibinfo {author} {\bibfnamefont
  {X.}~\bibnamefont {Xie}}, \bibinfo {author} {\bibfnamefont {T.~P.}\
  \bibnamefont {Devereaux}}, \bibinfo {author} {\bibfnamefont {M.~F.}\
  \bibnamefont {Crommie}}, \bibinfo {author} {\bibfnamefont {S.-K.}\
  \bibnamefont {Mo}},\ and\ \bibinfo {author} {\bibfnamefont {Z.-X.}\
  \bibnamefont {Shen}},\ }\bibfield  {title} {\bibinfo {title} {Quantum spin
  hall state in monolayer 1{T}’-{WT}e2},\ }\href
  {https://doi.org/10.1038/nphys4174} {\bibfield  {journal} {\bibinfo
  {journal} {Nature Physics}\ }\textbf {\bibinfo {volume} {13}},\ \bibinfo
  {pages} {683–687} (\bibinfo {year} {2017})}\BibitemShut {NoStop}%
\bibitem [{\citenamefont {Shi}\ \emph {et~al.}(2019)\citenamefont {Shi},
  \citenamefont {Kahn}, \citenamefont {Niu}, \citenamefont {Fei}, \citenamefont
  {Sun}, \citenamefont {Cai}, \citenamefont {Francisco}, \citenamefont {Wu},
  \citenamefont {Shen}, \citenamefont {Xu}, \citenamefont {Cobden},\ and\
  \citenamefont {Cui}}]{Shi2019}%
  \BibitemOpen
  \bibfield  {author} {\bibinfo {author} {\bibfnamefont {Y.}~\bibnamefont
  {Shi}}, \bibinfo {author} {\bibfnamefont {J.}~\bibnamefont {Kahn}}, \bibinfo
  {author} {\bibfnamefont {B.}~\bibnamefont {Niu}}, \bibinfo {author}
  {\bibfnamefont {Z.}~\bibnamefont {Fei}}, \bibinfo {author} {\bibfnamefont
  {B.}~\bibnamefont {Sun}}, \bibinfo {author} {\bibfnamefont {X.}~\bibnamefont
  {Cai}}, \bibinfo {author} {\bibfnamefont {B.~A.}\ \bibnamefont {Francisco}},
  \bibinfo {author} {\bibfnamefont {D.}~\bibnamefont {Wu}}, \bibinfo {author}
  {\bibfnamefont {Z.-X.}\ \bibnamefont {Shen}}, \bibinfo {author}
  {\bibfnamefont {X.}~\bibnamefont {Xu}}, \bibinfo {author} {\bibfnamefont
  {D.~H.}\ \bibnamefont {Cobden}},\ and\ \bibinfo {author} {\bibfnamefont
  {Y.-T.}\ \bibnamefont {Cui}},\ }\bibfield  {title} {\bibinfo {title} {Imaging
  quantum spin hall edges in monolayer {WT}e2},\ }\href
  {https://doi.org/10.1126/sciadv.aat8799} {\bibfield  {journal} {\bibinfo
  {journal} {Science Advances}\ }\textbf {\bibinfo {volume} {5}},\ \bibinfo
  {pages} {eaat8799} (\bibinfo {year} {2019})}\BibitemShut {NoStop}%
\bibitem [{\citenamefont {Garcia}\ \emph {et~al.}(2020)\citenamefont {Garcia},
  \citenamefont {Vila}, \citenamefont {Hsu}, \citenamefont {Waintal},
  \citenamefont {Pereira},\ and\ \citenamefont {Roche}}]{Garcia2020}%
  \BibitemOpen
  \bibfield  {author} {\bibinfo {author} {\bibfnamefont {J.~H.}\ \bibnamefont
  {Garcia}}, \bibinfo {author} {\bibfnamefont {M.}~\bibnamefont {Vila}},
  \bibinfo {author} {\bibfnamefont {C.-H.}\ \bibnamefont {Hsu}}, \bibinfo
  {author} {\bibfnamefont {X.}~\bibnamefont {Waintal}}, \bibinfo {author}
  {\bibfnamefont {V.~M.}\ \bibnamefont {Pereira}},\ and\ \bibinfo {author}
  {\bibfnamefont {S.}~\bibnamefont {Roche}},\ }\bibfield  {title} {\bibinfo
  {title} {Canted persistent spin texture and quantum spin hall effect
  {WT}e2},\ }\href {https://doi.org/10.1103/physrevlett.125.256603} {\bibfield
  {journal} {\bibinfo  {journal} {Physical Review Letters}\ }\textbf {\bibinfo
  {volume} {125}},\ \bibinfo {pages} {256603} (\bibinfo {year}
  {2020})}\BibitemShut {NoStop}%
\bibitem [{\citenamefont {Tang}\ \emph {et~al.}(2024)\citenamefont {Tang},
  \citenamefont {Ding}, \citenamefont {Chen}, \citenamefont {Gao},
  \citenamefont {Qian}, \citenamefont {Huang}, \citenamefont {Sun},
  \citenamefont {Han}, \citenamefont {Strasser}, \citenamefont {Li},
  \citenamefont {Geiwitz}, \citenamefont {Shehabeldin}, \citenamefont
  {Belosevich}, \citenamefont {Wang}, \citenamefont {Wang}, \citenamefont
  {Watanabe}, \citenamefont {Taniguchi}, \citenamefont {Bell}, \citenamefont
  {Wang}, \citenamefont {Fu}, \citenamefont {Zhang}, \citenamefont {Qian},
  \citenamefont {Burch}, \citenamefont {Shi}, \citenamefont {Ni}, \citenamefont
  {Chang}, \citenamefont {Xu},\ and\ \citenamefont {Ma}}]{Tang2024}%
  \BibitemOpen
  \bibfield  {author} {\bibinfo {author} {\bibfnamefont {J.}~\bibnamefont
  {Tang}}, \bibinfo {author} {\bibfnamefont {T.~S.}\ \bibnamefont {Ding}},
  \bibinfo {author} {\bibfnamefont {H.}~\bibnamefont {Chen}}, \bibinfo {author}
  {\bibfnamefont {A.}~\bibnamefont {Gao}}, \bibinfo {author} {\bibfnamefont
  {T.}~\bibnamefont {Qian}}, \bibinfo {author} {\bibfnamefont {Z.}~\bibnamefont
  {Huang}}, \bibinfo {author} {\bibfnamefont {Z.}~\bibnamefont {Sun}}, \bibinfo
  {author} {\bibfnamefont {X.}~\bibnamefont {Han}}, \bibinfo {author}
  {\bibfnamefont {A.}~\bibnamefont {Strasser}}, \bibinfo {author}
  {\bibfnamefont {J.}~\bibnamefont {Li}}, \bibinfo {author} {\bibfnamefont
  {M.}~\bibnamefont {Geiwitz}}, \bibinfo {author} {\bibfnamefont
  {M.}~\bibnamefont {Shehabeldin}}, \bibinfo {author} {\bibfnamefont
  {V.}~\bibnamefont {Belosevich}}, \bibinfo {author} {\bibfnamefont
  {Z.}~\bibnamefont {Wang}}, \bibinfo {author} {\bibfnamefont {Y.}~\bibnamefont
  {Wang}}, \bibinfo {author} {\bibfnamefont {K.}~\bibnamefont {Watanabe}},
  \bibinfo {author} {\bibfnamefont {T.}~\bibnamefont {Taniguchi}}, \bibinfo
  {author} {\bibfnamefont {D.~C.}\ \bibnamefont {Bell}}, \bibinfo {author}
  {\bibfnamefont {Z.}~\bibnamefont {Wang}}, \bibinfo {author} {\bibfnamefont
  {L.}~\bibnamefont {Fu}}, \bibinfo {author} {\bibfnamefont {Y.}~\bibnamefont
  {Zhang}}, \bibinfo {author} {\bibfnamefont {X.}~\bibnamefont {Qian}},
  \bibinfo {author} {\bibfnamefont {K.~S.}\ \bibnamefont {Burch}}, \bibinfo
  {author} {\bibfnamefont {Y.}~\bibnamefont {Shi}}, \bibinfo {author}
  {\bibfnamefont {N.}~\bibnamefont {Ni}}, \bibinfo {author} {\bibfnamefont
  {G.}~\bibnamefont {Chang}}, \bibinfo {author} {\bibfnamefont {S.-Y.}\
  \bibnamefont {Xu}},\ and\ \bibinfo {author} {\bibfnamefont {Q.}~\bibnamefont
  {Ma}},\ }\bibfield  {title} {\bibinfo {title} {Dual quantum spin hall
  insulator by density-tuned correlations in {T}a{I}r{T}e4},\ }\href
  {https://doi.org/10.1038/s41586-024-07211-8} {\bibfield  {journal} {\bibinfo
  {journal} {Nature}\ }\textbf {\bibinfo {volume} {628}},\ \bibinfo {pages}
  {515–521} (\bibinfo {year} {2024})}\BibitemShut {NoStop}%
\bibitem [{\citenamefont {Chen}\ \emph {et~al.}(2023)\citenamefont {Chen},
  \citenamefont {Wang}, \citenamefont {Li},\ and\ \citenamefont
  {Sanyal}}]{Chen2023}%
  \BibitemOpen
  \bibfield  {author} {\bibinfo {author} {\bibfnamefont {X.}~\bibnamefont
  {Chen}}, \bibinfo {author} {\bibfnamefont {D.}~\bibnamefont {Wang}}, \bibinfo
  {author} {\bibfnamefont {L.}~\bibnamefont {Li}},\ and\ \bibinfo {author}
  {\bibfnamefont {B.}~\bibnamefont {Sanyal}},\ }\bibfield  {title} {\bibinfo
  {title} {Giant spin-splitting and tunable spin-momentum locked transport in
  room temperature collinear antiferromagnetic semimetallic {C}r{O
  }monolayer},\ }\href {https://doi.org/10.1063/5.0147450} {\bibfield
  {journal} {\bibinfo  {journal} {Applied Physics Letters}\ }\textbf {\bibinfo
  {volume} {123}},\ \bibinfo {pages} {222404} (\bibinfo {year}
  {2023})}\BibitemShut {NoStop}%
\bibitem [{\citenamefont {Zou}\ \emph {et~al.}(2024)\citenamefont {Zou},
  \citenamefont {Yang}, \citenamefont {Xin}, \citenamefont {Wu}, \citenamefont
  {Cheng}, \citenamefont {Dong}, \citenamefont {Liu}, \citenamefont {Luo},
  \citenamefont {Lu},\ and\ \citenamefont {Wang}}]{Zou2024}%
  \BibitemOpen
  \bibfield  {author} {\bibinfo {author} {\bibfnamefont {K.}~\bibnamefont
  {Zou}}, \bibinfo {author} {\bibfnamefont {Y.}~\bibnamefont {Yang}}, \bibinfo
  {author} {\bibfnamefont {B.}~\bibnamefont {Xin}}, \bibinfo {author}
  {\bibfnamefont {W.}~\bibnamefont {Wu}}, \bibinfo {author} {\bibfnamefont
  {Y.}~\bibnamefont {Cheng}}, \bibinfo {author} {\bibfnamefont
  {H.}~\bibnamefont {Dong}}, \bibinfo {author} {\bibfnamefont {H.}~\bibnamefont
  {Liu}}, \bibinfo {author} {\bibfnamefont {F.}~\bibnamefont {Luo}}, \bibinfo
  {author} {\bibfnamefont {F.}~\bibnamefont {Lu}},\ and\ \bibinfo {author}
  {\bibfnamefont {W.-H.}\ \bibnamefont {Wang}},\ }\bibfield  {title} {\bibinfo
  {title} {Monolayer {M}2{X}2{O} as potential 2{D} altermagnets and
  half-metals: a first principles study},\ }\href
  {https://doi.org/10.1088/1361-648x/ad8e9f} {\bibfield  {journal} {\bibinfo
  {journal} {Journal of Physics: Condensed Matter}\ }\textbf {\bibinfo {volume}
  {37}},\ \bibinfo {pages} {055804} (\bibinfo {year} {2024})}\BibitemShut
  {NoStop}%
\bibitem [{\citenamefont {Goodenough}(1958)}]{Goodenough1958}%
  \BibitemOpen
  \bibfield  {author} {\bibinfo {author} {\bibfnamefont {J.~B.}\ \bibnamefont
  {Goodenough}},\ }\bibfield  {title} {\bibinfo {title} {An interpretation of
  the magnetic properties of the perovskite-type mixed crystals
  {L}a$_{1-x}${S}rx{C}o{O}$_{3-\lambda}$},\ }\href
  {https://doi.org/10.1016/0022-3697(58)90107-0} {\bibfield  {journal}
  {\bibinfo  {journal} {Journal of Physics and Chemistry of Solids}\ }\textbf
  {\bibinfo {volume} {6}},\ \bibinfo {pages} {287–297} (\bibinfo {year}
  {1958})}\BibitemShut {NoStop}%
\bibitem [{\citenamefont {Kanamori}(1959)}]{Kanamori1959}%
  \BibitemOpen
  \bibfield  {author} {\bibinfo {author} {\bibfnamefont {J.}~\bibnamefont
  {Kanamori}},\ }\bibfield  {title} {\bibinfo {title} {Superexchange
  interaction and symmetry properties of electron orbitals},\ }\href
  {https://doi.org/10.1016/0022-3697(59)90061-7} {\bibfield  {journal}
  {\bibinfo  {journal} {Journal of Physics and Chemistry of Solids}\ }\textbf
  {\bibinfo {volume} {10}},\ \bibinfo {pages} {87–98} (\bibinfo {year}
  {1959})}\BibitemShut {NoStop}%
\bibitem [{\citenamefont {Anderson}\ and\ \citenamefont
  {Hasegawa}(1955)}]{Anderson1955}%
  \BibitemOpen
  \bibfield  {author} {\bibinfo {author} {\bibfnamefont {P.~W.}\ \bibnamefont
  {Anderson}}\ and\ \bibinfo {author} {\bibfnamefont {H.}~\bibnamefont
  {Hasegawa}},\ }\bibfield  {title} {\bibinfo {title} {Considerations on double
  exchange},\ }\href {https://doi.org/10.1103/physrev.100.675} {\bibfield
  {journal} {\bibinfo  {journal} {Physical Review}\ }\textbf {\bibinfo {volume}
  {100}},\ \bibinfo {pages} {675–681} (\bibinfo {year} {1955})}\BibitemShut
  {NoStop}%
\bibitem [{\citenamefont {Wang}\ \emph {et~al.}(2015)\citenamefont {Wang},
  \citenamefont {Deng}, \citenamefont {Liu},\ and\ \citenamefont
  {Liu}}]{Wang2015}%
  \BibitemOpen
  \bibfield  {author} {\bibinfo {author} {\bibfnamefont {J.}~\bibnamefont
  {Wang}}, \bibinfo {author} {\bibfnamefont {S.}~\bibnamefont {Deng}}, \bibinfo
  {author} {\bibfnamefont {Z.}~\bibnamefont {Liu}},\ and\ \bibinfo {author}
  {\bibfnamefont {Z.}~\bibnamefont {Liu}},\ }\bibfield  {title} {\bibinfo
  {title} {The rare two-dimensional materials with dirac cones},\ }\href
  {https://doi.org/10.1093/nsr/nwu080} {\bibfield  {journal} {\bibinfo
  {journal} {National Science Review}\ }\textbf {\bibinfo {volume} {2}},\
  \bibinfo {pages} {22–39} (\bibinfo {year} {2015})}\BibitemShut {NoStop}%
\bibitem [{\citenamefont {Kresse}\ and\ \citenamefont
  {Furthmüller}(1996)}]{KRESSE199615}%
  \BibitemOpen
  \bibfield  {author} {\bibinfo {author} {\bibfnamefont {G.}~\bibnamefont
  {Kresse}}\ and\ \bibinfo {author} {\bibfnamefont {J.}~\bibnamefont
  {Furthmüller}},\ }\bibfield  {title} {\bibinfo {title} {Efficiency of
  ab-initio total energy calculations for metals and semiconductors using a
  plane-wave basis set},\ }\href
  {https://doi.org/https://doi.org/10.1016/0927-0256(96)00008-0} {\bibfield
  {journal} {\bibinfo  {journal} {Computational Materials Science}\ }\textbf
  {\bibinfo {volume} {6}},\ \bibinfo {pages} {15} (\bibinfo {year}
  {1996})}\BibitemShut {NoStop}%
\bibitem [{\citenamefont {Kresse}\ and\ \citenamefont
  {Furthm\"uller}(1996)}]{PhysRevB.54.11169}%
  \BibitemOpen
  \bibfield  {author} {\bibinfo {author} {\bibfnamefont {G.}~\bibnamefont
  {Kresse}}\ and\ \bibinfo {author} {\bibfnamefont {J.}~\bibnamefont
  {Furthm\"uller}},\ }\bibfield  {title} {\bibinfo {title} {Efficient iterative
  schemes for ab initio total-energy calculations using a plane-wave basis
  set},\ }\href {https://doi.org/10.1103/PhysRevB.54.11169} {\bibfield
  {journal} {\bibinfo  {journal} {Physical Review B}\ }\textbf {\bibinfo
  {volume} {54}},\ \bibinfo {pages} {11169} (\bibinfo {year}
  {1996})}\BibitemShut {NoStop}%
\bibitem [{\citenamefont {Perdew}\ \emph {et~al.}(1996)\citenamefont {Perdew},
  \citenamefont {Burke},\ and\ \citenamefont
  {Ernzerhof}}]{PhysRevLett.77.3865}%
  \BibitemOpen
  \bibfield  {author} {\bibinfo {author} {\bibfnamefont {J.~P.}\ \bibnamefont
  {Perdew}}, \bibinfo {author} {\bibfnamefont {K.}~\bibnamefont {Burke}},\ and\
  \bibinfo {author} {\bibfnamefont {M.}~\bibnamefont {Ernzerhof}},\ }\bibfield
  {title} {\bibinfo {title} {Generalized gradient approximation made simple},\
  }\href {https://doi.org/10.1103/PhysRevLett.77.3865} {\bibfield  {journal}
  {\bibinfo  {journal} {Physical Review Letters}\ }\textbf {\bibinfo {volume}
  {77}},\ \bibinfo {pages} {3865} (\bibinfo {year} {1996})}\BibitemShut
  {NoStop}%
\bibitem [{\citenamefont {Bl\"ochl}(1994)}]{PhysRevB.50.17953}%
  \BibitemOpen
  \bibfield  {author} {\bibinfo {author} {\bibfnamefont {P.~E.}\ \bibnamefont
  {Bl\"ochl}},\ }\bibfield  {title} {\bibinfo {title} {Projector augmented-wave
  method},\ }\href {https://doi.org/10.1103/PhysRevB.50.17953} {\bibfield
  {journal} {\bibinfo  {journal} {Physical Review B}\ }\textbf {\bibinfo
  {volume} {50}},\ \bibinfo {pages} {17953} (\bibinfo {year}
  {1994})}\BibitemShut {NoStop}%
\bibitem [{\citenamefont {Kresse}\ and\ \citenamefont
  {Joubert}(1999)}]{PhysRevB.59.1758}%
  \BibitemOpen
  \bibfield  {author} {\bibinfo {author} {\bibfnamefont {G.}~\bibnamefont
  {Kresse}}\ and\ \bibinfo {author} {\bibfnamefont {D.}~\bibnamefont
  {Joubert}},\ }\bibfield  {title} {\bibinfo {title} {From ultrasoft
  pseudopotentials to the projector augmented-wave method},\ }\href
  {https://doi.org/10.1103/PhysRevB.59.1758} {\bibfield  {journal} {\bibinfo
  {journal} {Physical Review B}\ }\textbf {\bibinfo {volume} {59}},\ \bibinfo
  {pages} {1758} (\bibinfo {year} {1999})}\BibitemShut {NoStop}%
\bibitem [{\citenamefont {Wisesa}\ \emph {et~al.}(2016)\citenamefont {Wisesa},
  \citenamefont {McGill},\ and\ \citenamefont {Mueller}}]{PhysRevB.93.155109}%
  \BibitemOpen
  \bibfield  {author} {\bibinfo {author} {\bibfnamefont {P.}~\bibnamefont
  {Wisesa}}, \bibinfo {author} {\bibfnamefont {K.~A.}\ \bibnamefont {McGill}},\
  and\ \bibinfo {author} {\bibfnamefont {T.}~\bibnamefont {Mueller}},\
  }\bibfield  {title} {\bibinfo {title} {Efficient generation of generalized
  monkhorst-pack grids through the use of informatics},\ }\href
  {https://doi.org/10.1103/PhysRevB.93.155109} {\bibfield  {journal} {\bibinfo
  {journal} {Physical Review B}\ }\textbf {\bibinfo {volume} {93}},\ \bibinfo
  {pages} {155109} (\bibinfo {year} {2016})}\BibitemShut {NoStop}%
\bibitem [{\citenamefont {Moore}\ \emph {et~al.}(2024)\citenamefont {Moore},
  \citenamefont {Horton}, \citenamefont {Linscott}, \citenamefont {Ganose},
  \citenamefont {Siron}, \citenamefont {O'Regan},\ and\ \citenamefont
  {Persson}}]{DFTU_1}%
  \BibitemOpen
  \bibfield  {author} {\bibinfo {author} {\bibfnamefont {G.~C.}\ \bibnamefont
  {Moore}}, \bibinfo {author} {\bibfnamefont {M.~K.}\ \bibnamefont {Horton}},
  \bibinfo {author} {\bibfnamefont {E.}~\bibnamefont {Linscott}}, \bibinfo
  {author} {\bibfnamefont {A.~M.}\ \bibnamefont {Ganose}}, \bibinfo {author}
  {\bibfnamefont {M.}~\bibnamefont {Siron}}, \bibinfo {author} {\bibfnamefont
  {D.~D.}\ \bibnamefont {O'Regan}},\ and\ \bibinfo {author} {\bibfnamefont
  {K.~A.}\ \bibnamefont {Persson}},\ }\bibfield  {title} {\bibinfo {title}
  {High-throughput determination of hubbard {U} and hund {J} values for
  transition metal oxides via the linear response formalism},\ }\href
  {https://doi.org/10.1103/PhysRevMaterials.8.014409} {\bibfield  {journal}
  {\bibinfo  {journal} {Physical Review Materials}\ }\textbf {\bibinfo {volume}
  {8}},\ \bibinfo {pages} {014409} (\bibinfo {year} {2024})}\BibitemShut
  {NoStop}%
\bibitem [{\citenamefont {Baroni}\ \emph {et~al.}(2001)\citenamefont {Baroni},
  \citenamefont {de~Gironcoli}, \citenamefont {Dal~Corso},\ and\ \citenamefont
  {Giannozzi}}]{baroni2001}%
  \BibitemOpen
  \bibfield  {author} {\bibinfo {author} {\bibfnamefont {S.}~\bibnamefont
  {Baroni}}, \bibinfo {author} {\bibfnamefont {S.}~\bibnamefont
  {de~Gironcoli}}, \bibinfo {author} {\bibfnamefont {A.}~\bibnamefont
  {Dal~Corso}},\ and\ \bibinfo {author} {\bibfnamefont {P.}~\bibnamefont
  {Giannozzi}},\ }\bibfield  {title} {\bibinfo {title} {Phonons and related
  crystal properties from density-functional perturbation theory},\ }\href
  {https://doi.org/10.1103/RevModPhys.73.515} {\bibfield  {journal} {\bibinfo
  {journal} {Reviews of Modern Physics}\ }\textbf {\bibinfo {volume} {73}},\
  \bibinfo {pages} {515} (\bibinfo {year} {2001})}\BibitemShut {NoStop}%
\bibitem [{\citenamefont {Togo}\ \emph {et~al.}(2023)\citenamefont {Togo},
  \citenamefont {Chaput}, \citenamefont {Tadano},\ and\ \citenamefont
  {Tanaka}}]{phonopy-phono3py-JPCM}%
  \BibitemOpen
  \bibfield  {author} {\bibinfo {author} {\bibfnamefont {A.}~\bibnamefont
  {Togo}}, \bibinfo {author} {\bibfnamefont {L.}~\bibnamefont {Chaput}},
  \bibinfo {author} {\bibfnamefont {T.}~\bibnamefont {Tadano}},\ and\ \bibinfo
  {author} {\bibfnamefont {I.}~\bibnamefont {Tanaka}},\ }\bibfield  {title}
  {\bibinfo {title} {Implementation strategies in phonopy and phono3py},\
  }\href {https://doi.org/10.1088/1361-648X/acd831} {\bibfield  {journal}
  {\bibinfo  {journal} {Journal of Physics: Condensed Matter}\ }\textbf
  {\bibinfo {volume} {35}},\ \bibinfo {pages} {353001} (\bibinfo {year}
  {2023})}\BibitemShut {NoStop}%
\bibitem [{\citenamefont {Togo}(2023)}]{phonopy-phono3py-JPSJ}%
  \BibitemOpen
  \bibfield  {author} {\bibinfo {author} {\bibfnamefont {A.}~\bibnamefont
  {Togo}},\ }\bibfield  {title} {\bibinfo {title} {First-principles phonon
  calculations with phonopy and phono3py},\ }\href
  {https://doi.org/10.7566/JPSJ.92.012001} {\bibfield  {journal} {\bibinfo
  {journal} {Journal of the Physical Society of Japan}\ }\textbf {\bibinfo
  {volume} {92}},\ \bibinfo {pages} {012001} (\bibinfo {year}
  {2023})}\BibitemShut {NoStop}%
\bibitem [{\citenamefont {Esfarjani}\ and\ \citenamefont
  {Stokes}(2008)}]{PhysRevB.77.144112}%
  \BibitemOpen
  \bibfield  {author} {\bibinfo {author} {\bibfnamefont {K.}~\bibnamefont
  {Esfarjani}}\ and\ \bibinfo {author} {\bibfnamefont {H.~T.}\ \bibnamefont
  {Stokes}},\ }\bibfield  {title} {\bibinfo {title} {Method to extract
  anharmonic force constants from first principles calculations},\ }\href
  {https://doi.org/10.1103/PhysRevB.77.144112} {\bibfield  {journal} {\bibinfo
  {journal} {Phys. Rev. B}\ }\textbf {\bibinfo {volume} {77}},\ \bibinfo
  {pages} {144112} (\bibinfo {year} {2008})}\BibitemShut {NoStop}%
\bibitem [{\citenamefont {Hellman}\ \emph {et~al.}(2011)\citenamefont
  {Hellman}, \citenamefont {Abrikosov},\ and\ \citenamefont
  {Simak}}]{PhysRevB.84.180301}%
  \BibitemOpen
  \bibfield  {author} {\bibinfo {author} {\bibfnamefont {O.}~\bibnamefont
  {Hellman}}, \bibinfo {author} {\bibfnamefont {I.~A.}\ \bibnamefont
  {Abrikosov}},\ and\ \bibinfo {author} {\bibfnamefont {S.~I.}\ \bibnamefont
  {Simak}},\ }\bibfield  {title} {\bibinfo {title} {Lattice dynamics of
  anharmonic solids from first principles},\ }\href
  {https://doi.org/10.1103/PhysRevB.84.180301} {\bibfield  {journal} {\bibinfo
  {journal} {Phys. Rev. B}\ }\textbf {\bibinfo {volume} {84}},\ \bibinfo
  {pages} {180301} (\bibinfo {year} {2011})}\BibitemShut {NoStop}%
\bibitem [{\citenamefont {Nosé}(1984)}]{Nos1984}%
  \BibitemOpen
  \bibfield  {author} {\bibinfo {author} {\bibfnamefont {S.}~\bibnamefont
  {Nosé}},\ }\bibfield  {title} {\bibinfo {title} {A unified formulation of
  the constant temperature molecular dynamics methods},\ }\href
  {https://doi.org/10.1063/1.447334} {\bibfield  {journal} {\bibinfo  {journal}
  {The Journal of Chemical Physics}\ }\textbf {\bibinfo {volume} {81}},\
  \bibinfo {pages} {511–519} (\bibinfo {year} {1984})}\BibitemShut {NoStop}%
\bibitem [{\citenamefont {Nosé}(1991)}]{Nos1991}%
  \BibitemOpen
  \bibfield  {author} {\bibinfo {author} {\bibfnamefont {S.}~\bibnamefont
  {Nosé}},\ }\bibfield  {title} {\bibinfo {title} {Constant temperature
  molecular dynamics methods},\ }\href {https://doi.org/10.1143/ptps.103.1}
  {\bibfield  {journal} {\bibinfo  {journal} {Progress of Theoretical Physics
  Supplement}\ }\textbf {\bibinfo {volume} {103}},\ \bibinfo {pages} {1–46}
  (\bibinfo {year} {1991})}\BibitemShut {NoStop}%
\bibitem [{\citenamefont {Hoover}(1985)}]{Hoover1985}%
  \BibitemOpen
  \bibfield  {author} {\bibinfo {author} {\bibfnamefont {W.~G.}\ \bibnamefont
  {Hoover}},\ }\bibfield  {title} {\bibinfo {title} {Canonical dynamics:
  Equilibrium phase-space distributions},\ }\href
  {https://doi.org/10.1103/physreva.31.1695} {\bibfield  {journal} {\bibinfo
  {journal} {Physical Review A}\ }\textbf {\bibinfo {volume} {31}},\ \bibinfo
  {pages} {1695–1697} (\bibinfo {year} {1985})}\BibitemShut {NoStop}%
\end{thebibliography}%





\end{document}